\documentclass[
 twocolumn%
 ,secnumarabic%
,tightenlines%
,amssymb, amsmath,nobibnotes, aps, prb,showpacs]{revtex4}
\usepackage{docs}%
\usepackage{amsfonts,amssymb,amsmath,amsopn}
\usepackage{bm}
\usepackage{isolatin1}
\usepackage{graphicx}
\usepackage[german]{varioref}
\usepackage[dvips]{epsfig}
\usepackage{graphics}
\usepackage[dvips]{color}
\usepackage{color}
\usepackage{colordvi}
\usepackage{dsfont}
\usepackage{psfrag}
\expandafter\ifx\csname package@font\endcsname\relax\else
 \expandafter\expandafter
 \expandafter\usepackage
 \expandafter\expandafter
 \expandafter{\csname package@font\endcsname}%
\fi
\begin{document}
\newcommand{\captionfonts}{\small}
\newcommand{\be}{\begin{eqnarray}}
\newcommand{\ee}{\end{eqnarray}}
\def\p#1#2{|#1\rangle \langle #2|}
\def\ket#1{|#1\rangle}
\def\bra#1{\langle #1|}
\def\refeq#1{(\ref{#1})}
\def\tb#1{{\overline{{\underline{ #1}}}}}
\def\im{\mbox{Im}}
\def\re{\mbox{Re}}
\def\nn{\nonumber}
\def\t{\mbox{tr}}
\def\sgn{\mbox{sgn}}
\def\Li{\mbox{Li}}
\def\P{\mbox{P}}
\def\d{\mbox d}
\def\i{\int_{-\infty}^{\infty}}
\def\ip{\int_{0}^{\infty}}
\def\mi{\int_{-\infty}^{0}}
\def\A{\mathfrak A}
\def\AA{{\overline{{\mathfrak{A}}}}}
\def\a{\mathfrak a}
\def\aa{{\overline{{\mathfrak{a}}}}}
\def\B{\mathfrak B}
\def\BB{{\overline{{\mathfrak{B}}}}}
\def\b{\mathfrak b}
\def\bb{{\overline{{\mathfrak{b}}}}}
\def\R{\mathcal R}
\def\dm{\mathfrak d}
\def\dd{{\overline{{\mathfrak{d}}}}}
\def\D{\mathfrak D}
\def\DD{{\overline{{\mathfrak{D}}}}}
\def\c{\mathfrak c}
\def\cc{{\overline{{\mathfrak{c}}}}}
\def\C{\mathfrak C}
\def\CC{{\overline{{\mathfrak{C}}}}}
\def\O{\mathcal O}
\def\F{\mathcal F_k}
\def\N{\mathcal N}
\def\I{\mathcal I}
\def\S{\mathcal S}
\def\G{\Gamma}
\def\L{\Lambda}
\def\la{\lambda}
\def\g{\gamma}
\def\al{\alpha}
\def\s{\sigma}
\def\e{\epsilon}
\def\k{\kappa}
\def\ve{\varepsilon}
\def\te{\text{e}}
\def\ti{\text{i}}
\def\max{\text{max}}
\def\str{\text{str}}
\def\tr{\text{tr}}
\def\tC{\text C}
\def\Fo{\mathcal{F}_{1,k}}
\def\Ft{\mathcal{F}_{2,k}}
\def\vs{\varsigma}
\def\l{\left}
\def\r{\right}
\def\up{\uparrow}
\def\down{\downarrow}
\def\u{\underline}
\def\ov{\overline}
\title{The anisotropic multichannel spin-$S$ Kondo model:\\ Calculation of scales from a novel exact solution}
\author{Michael Bortz, Andreas Kl\"umper}%
%
%
\affiliation{Bergische Universit\"at Wuppertal, Fachbereich Physik, 42097 Wuppertal, Germany}
\date{June 9, 2004}%
%
\begin{abstract}
A novel exact solution of the multichannel spin-$S$ Kondo model is presented, based on a lattice path integral approach of the single channel spin-$1/2$ case. The spin exchange between the localized moment and the host is of $XXZ$-type, including the isotropic $XXX$ limit. The free energy is given by a finite set of non-linear integral equations, which allow for an accurate determination of high- and low-temperature scales.
\end{abstract}
\maketitle

%


\section{Introduction}
The isotropic multichannel spin-$S$ Kondo model describes the $XXX$-like spin
scattering between non-interacting spin-1/2 fermions in $m$ channels and a
single localized spin-$S$ impurity. It has been proposed by Nozi\`eres and
Blandin\cite{noz80} to account for the orbital structure of spin-$S$
impurities in metals. Physical realizations are discussed in the review
article \cite{sch93}. An exact solution to this model was found by
Tsvelick and Wiegmann \cite{tsv83a,tsv84, wie83} using the Bethe Ansatz
(BA). By applying the thermodynamic Bethe Ansatz (TBA), the thermodynamics has
been calculated by Tsvelick \cite{tsv85} and Andrei and Destri \cite{and84}. In the TBA approach, the free energy
is encoded by infinitely many coupled non-linear integral equations. Recently,
Schlottmann \cite{sch00, sch01} and Zar\'and et al \cite{zar02} obtained the
free energy for the anisotropic multichannel spin-$S$ Kondo model with
$XXZ$-like exchange in the TBA approach. Schlottmann observed a quantum
critical point for the overscreened model in the limit of low temperatures, Zar\'and et al focused on the underscreened two-channel model. Here, we present a novel exact solution in which the free energy is encoded in a {\em finite} set of non-linear integral equations (NLIE), including both the anisotropic and isotropic models for arbitrary $S$ and $m$. On the one hand, we confirm known results by this novel solution, on the other hand, we obtain farther reaching results as far as ratios of high- and low-temperature scales are concerned. 

In a recent publication \cite{bo04}, we proposed a lattice path integral approach to an Anderson-like impurity in a correlated host. We obtained this model from the Hamiltonian limit of a gl(2$|$1) symmetric transfer matrix. The free energy contributions of both the host and the impurity were calculated exactly by combining the Trotter Suzuki mapping with the quantum transfer matrix technique. We showed that the model allows for the Kondo limit, i.e. a localized magnetic impurity in a free host with linearized energy-momentum relation. In the Kondo limit, the symmetry of the coupling between host and impurity is reduced from gl(2$|$1) to su(2). In the following, we exploit this observation to propose a U$_q$su(2) symmetric quantum transfer matrix (QTM), whose largest eigenvalue yields the free energy contribution of the impurity in the thermodynamic limit, in close analogy to \cite{bo04}. The $q$-deformation accounts for an $XXZ$-like spin exchange between the localized moment and the host, including the isotropic limit $q\to 1$. The transfer matrix is built from local $R$-operators, each acting in the tensor product space of two U$_q$su(2) modules, carrying an $(m+1)$-dimensional irrep and an $(l+1)$-dimensional irrep of U$_q$su(2), respectively. Here $m$ denotes the number of the electronic host channels and $l=2S$ is twice the impurity spin. By making use of analyticity arguments, the free energy is represented by $($max$[l,m]+1)$-many NLIE. These equations are investigated both analytically and numerically for the accurate calculation of scales in the limits of low and high temperatures. 

This article is organized as follows. In the next section, the free energy
contribution of the impurity is determined as the largest eigenvalue of a
certain QTM. The third and fourth sections are devoted to the calculation of thermodynamic equilibrium functions and scales in the limits of
low and high temperatures, respectively. In the last section, we summarize the results. 

In all that follows we set $k_B=1$, and $g\mu_B=1$, where $k_B$ is Boltzmann's
constant, $g$ is the gyromagnetic factor and $\mu_B$ is the Bohr magneton. An
index $i$ ($h$) denotes quantities pertaining to the impurity (host). 

\section{Calculation of the free energy}
Let $V_q^{(l)}$ be the module carrying the $l$-dimensional irrep of
U$_q$su(2); in the limit $q=1$, $V_1^{(l)}$ carries the $l$-dimensional irrep
of su(2). Consider the matrices $R_q^{(l,m)}(x)\in$End$\l(V^{(l)}_q\otimes
V^{(m)}_q\r)$ constructed such that the Yang-Baxter-Equation (YBE) is fulfilled:
\be
\lefteqn{\l[R_q^{(l,m)}(x)\r]^{\beta, \g}_{\beta', \g'} \l[R_q^{(n,m)}(y)\r]^{\alpha, \g'}_{\alpha', \g''}
\l[R_q^{(n,l)}(y-x)\r]^{\alpha', \beta'}_{\alpha'',
  \beta''}}\nn\\
& &\!\!\!\!\!=\!\!\l[R_q^{(n,l)}(y-x)\r]^{\alpha, \beta}_{\alpha',\beta'} \l[R_q^{(n,m)}(y)\r]^{\alpha', \g}_{\alpha'', \g'}
\l[R_q^{(l,m)}(x)\r]^{\beta', \g'}_{\beta'',
  \g''}\nn\, .
\ee
The $R_q^{(l,m)}(x)$ are obtained by fusing a
lattice of $m\times l$ many operators $R_q^{(1,1)}$. Then $V_q^{(k)}$ is the subspace of completely symmetric tensors in $\l[V_{q}^{(1)}\r]^{\otimes k}$, with $k=l,m$. In \cite{kir86}, the explicit expression of $R_q^{(l,m)}(u)$ is given. 
If in the isotropic single channel case, the spectral parameter is chosen to take the special value $x_0:=\l[\frac{2}{l+1}
\tan J\l(l+1\r)\r]^{-1}$, the matrix $R^{(l,1)}_1(x_0)$ simplifies to \cite{tsv83}
\be
  R^{(l,1)}_1(x_0)&=& \exp\l[\ti2 J {\mathbf S}\cdot  \mbox{\boldmath $\sigma$}\r]\label{spinsm1}\;,
\ee
The logarithm of equation \refeq{spinsm1} is the
spin exchange operator between the impurity spin $S=l/2$ and one electron, where the spin-exchange coupling $J$ is parameterized by $x_0$. 

It has been argued by Tsvelick et al \cite{tsv83a} that the exchange in the multichannel Kondo model is equivalent to an exchange model for spin-$m/2$ electrons scattered by the spin-$S$ impurity. This equivalence has been proven by the equality of the exact solutions of both models at zero temperature. We expect that more insight into this question is possible by a lattice path integral formulation in
analogy to the $S=1/2,\,m=1,\,\g=0$ case \cite{bo04}. We leave
this task as a future challenge. Here, we rely on the above mentioned equivalence, so that the Hamiltonian density is given by: 
\begin{subequations}
\be
\mathcal H\!\!\!&=&\!\!\!\mathcal H_h+\mathcal H_{sd}+\mathcal H_{ex}\nn\\
\mathcal H_h\!\!\!&=&\!\!\!-\ti  v_f\sum_{\sigma=\pm 1/2}\sum_{k=1}^{m}
\text{:}\psi^\dagger_{\sigma,k}(x)\frac{\d}{\d x}\psi_{\sigma,k}(x)\text{:}\label{defhhgen}\\
\mathcal H_{i}\!\!\!&=&\!\!\!
-\ti\delta(x)\sum_{\sigma,\sigma'}\text{:}\psi^\dagger_{\sigma}(x)\l[\ln  
R_q^{(l,m)}(x_0)\r]_\sigma^{\sigma'}\psi_{\sigma'}(x)\text{:}\label{defhsdgen}\\
\mathcal H_{ex}\!\!\!&=&\!\!\!\frac h2\l[\sum_{\sigma=-m/2}^{m/2}\sigma n_{\sigma}(x)+\delta(x)\sum_{\tau=-S}^S\tau
n_{d,\tau}\r]\label{defhexgen}\\
n_{\sigma,k}\!\!\!&:=&\!\!\!\text{:}\psi^\dagger_{\sigma,k}\psi_{\sigma,k}\text{:}\;,\;n_{k}=\sum_\sigma n_{\sigma,k}\;,\; n_{\sigma}=\sum_k
n_{\sigma,k}\nn\;,
\ee
\end{subequations}
where $x_0$ and $q$ parameterize the coupling constants. In the following, $q$ is given by a real constant $\g$, $q=:\exp(\ti\g)$, such that $\g\neq 0$ induces an Ising-like anisotropy, $J_\perp<J_{||}$. The matrix $R_q^{(l,m)}$ acts in the electronic space with
operator-valued entries in the impurity space. Only the totally symmetric combination of spinors, with total spin $m/2$, interacts with the impurity, so that the spin indices in equation \refeq{defhsdgen} are $\sigma,\sigma'=-m/2,\ldots,m/2$. 

Consider the monodromy matrix
\be
T_q^{(l,m)}\!\!\!&:=&\!\!\! \l[R_q^{(l,m)}\r]_{a,N} \l[R_q^{(l,m)}\r]_{a,N-1}\ldots
 \l[R_q^{(l,m)}\r]_{a,1}\label{deftlmeff}\\
\tau_q^{(l,m)}\!\!\!&=&\!\!\!\t_a T^{(l,m)}\nn\; ,
\ee
where $\l[R_q^{(l,m)}\r]_{a,n}$ acts in the tensor product $V^{(l)}_a\otimes V^{(m)}_n$. In the last line the transfer matrix $\tau_q^{(l,m)}$ is defined as the
trace over the auxiliary space of $T_q^{(l,m)}$. In view of the definition of
the Hamiltonian in equation \refeq{defhsdgen}, the auxiliary space $a$ is identified with the impurity space and the quantum spaces $1,\ldots,N$ correspond to the
electrons. We will now show that under the substitution $N\to \ti
\beta D$, the largest eigenvalue $\L_1^{(1,1)}(x_0)$ of $T_1^{(1,1)}(x_0)$ is directly related to the
free energy contribution of the impurity $f_i$ found in \cite{bo04},
\be
-\beta f_i= \l.\ln \L_1^{(1,1)}(x_0)\r|_{N\to\ti\beta D}\label{sinchan}\; .
\ee
Here $D$ has the meaning of a bandwidth parameter of the host. A
magnetic field is introduced by twisted boundary conditions. 

The largest eigenvalue in equation \refeq{sinchan} is expressed through auxiliary
functions $\b,\bb$ ($\B=1+\b$, $\BB=1+\bb$) in analogy to \cite{kl93}. After the substitution $N\to\ti\beta D$, these functions are given by the non-linear integral equation
\be
\ln \b(x)&=&-2D\beta \arctan \te^x +\frac{\beta h}{2} \nn\\
& &+\l[\k*\ln \B-\k_-*\ln
  \BB\r](x)\label{s1m1iso}\;,
\ee
where $\k(x):= \frac{1}{2\pi}\i \frac{\te^{-\pi/2|k|}}{2\cosh\frac{\pi
    k}{2}}\te^{\ti kx}\d k$ and $\k_\pm(x):=\k(x\pm\ti\pi)$. An equation similar
to \refeq{s1m1iso} holds with $\b\to \bb$, $h\to -h$ and $\k_-\to \k_+$. The free energy is obtained from the auxiliary functions by
\be
-\beta f_i&=& \frac{1}{2\pi}\int_{-\infty}^\infty
\frac{1}{\cosh(x+x_0)}\ln[\B\BB](x)\d x\nn\; .
\ee
These equations are equivalent to the corresponding TBA equations
\cite{tsv83}, as can be shown by fusion techniques described in \cite{jue97eq}. Proceeding as in
\cite{tsv83}, the universal limit $D\to\infty$ is taken by defining the temperature
scale 
\be
T_K:= 2D \te^{-\pi x_0}\; .\label{tkdef}
\ee
This results in the $D$-independent equations
\be
\ln \b(x)&=&- \te^x +\frac{\beta h}{2} +\l[\k*\ln \B-\k_-*\ln
  \BB\r](x)\nn\\
-\beta f_i&=& \frac{1}{2\pi}\int_{-\infty}^\infty
\frac{1}{\cosh(x+T/T_K)}\ln[\B\BB](x)\d x\nn\; ,
\ee 
in complete agreement with results from \cite{bo04}. Thus we argue that the
impurity contribution to the free energy of the anisotropic multichannel
spin-$S$ Kondo model is given by
\be
-\beta f_i=\l.\ln \L_q^{(l,m)}(x_0)\r|_{N\to\ti\beta D}\label{mulchan}\;,
\ee
where the universal limit $D\to\infty$ is implicit. 

The eigenvalue $\L_q^{(l,m)}$ is found by standard Bethe-Ansatz techniques \cite{kir86,bab83}. 
Following \cite{suz98}, one introduces suitably chosen
auxiliary functions in order to exploit the analyticity properties of the
largest eigenvalue. We restrict ourselves to the anisotropy parameter
\be
\g<\frac{\pi}{2 n}=: \g_{\text{max}}\label{restan}\; ,
\ee
where $n:= \max(l,m)$. We define a temperature scale
\be
T_K&=& 2 D\te^{-\pi x_0/\g}\label{tkeffdef}\; ,
\ee
from which the isotropic case \refeq{tkdef} is recovered by scaling $x_0\to \g x_0$. One is left with the following system of NLIE for the auxiliary functions $y_j=1+Y_j$, $(j=1,\ldots,n-1)$, $\b_n=1+\B_n$ and $\bb_n=1+\BB_n$:
\be
{\mathbf y}(x)\!\!\! &=& \!\!\! {\mathbf d}(x) + [\hat{\mathbf  \k}* {\mathbf Y}](x)\label{nlis} \\
{\mathbf y}\!\!\! &:=&\!\!\! \left( \ln y_1 , \ln y_2, \cdots , \ln y_{n-1}, \ln \b_n, \ln \bb_n\right)^T\nn \\
{\mathbf Y}\!\!\! &:=&\!\!\! \left(\ln Y_1,\ln Y_2 ,\cdots,\ln \B_n, \ln \BB_n\right)^T\nn \\
{\mathbf d}(x)\!\!\! &:=&\!\!\! \left( 0,0,\ldots,0,\underbrace{-\te^x}_{m^{\text{th}}\text{entry}},0,\ldots, 0,f_\g\beta h ,-f_\g\beta h\right)^T \nn \\
\hat{\mathbf  \k}(x)\!\!\! &=&\!\!\! \left( \begin{array}{ccccc}
0     &s(x)  &   0     &   \cdots     &0\\
s(x)  &0     & s(x)    &0                 &\vdots\\
0     &\ddots&  \ddots &    \ddots        &      0         \\
0     &\ldots&  0           &s(x)      &s(x)\\
0     &\ldots& s(x)&\k(x)              &-\k_-(x)\\ 
0     &\ldots& s(x)&-\k_+(x)&\k(x)
\end{array}\right) \label{matvec}\\
\k(x)\!\!\!&=&\!\!\! \frac{1}{2\pi} \i\frac{\sinh\frac{\pi}{2}
  k\l(\frac{\pi}{\g}-(n+1)\r)}{2\cosh\frac{\pi k}{2}\,\sinh\frac{\pi }{2}k\l(\frac{\pi}{\g}-n\r)}\, \te^{\ti kx}\d k\label{intkdef}\\
\!\!\!&\stackrel{\g\to0}{=}&\!\!\!\frac{1}{2\pi } \, \i \, \frac{\te^{-\frac{\pi}{2}\,
  |k|}}{2\cosh \frac{\pi k}{2}}\, \te^{\ti kx}\,\d k\nn \\
s(x)\!\!\!&=&\!\!\! \frac{1}{2\, \cosh  x}\,,\; f_\g= \frac{1}{2(1-n\g/\pi)}\stackrel{\g\to0}{=} \frac{1}{2}\label{restdef}\; , 
\ee
and the free energy contribution of the impurity is given by
\be
f_i(T,h)=-T\l\{\begin{array}{ll}
   \i\frac{\ln Y_{l}(x)}{2\pi \cosh\l(x+\ln \frac{T}{T_K}\r)}\,\d
  x\,,&l<m \\
 \i\frac{\ln\l[ \B_l\BB_{l}\r](x)}{2\pi \cosh\l(x+\ln \frac{T}{T_K}\r)}\,\d
  x\,,&l\geq m \end{array}\r. \label{mchsspev}\; .
\ee
Again, the magnetic field is introduced by twisted boundary conditions. The integration kernel $\k$ depends on $n$ if
  $\g\neq0$. The asymmetric driving term $-\te^{x}$ in the $m$-th equation
  gives rise to different asymptotes in the limits $x\to\infty$,
  $x\to-\infty$, summarized in the following:  
\begin{subequations} 
\label{asympval}
\be
\ln Y_{j}^{(-\infty)}&=&2\ln\frac{\sinh\frac{\beta h}{2}(j+1)}{\sinh\frac{\beta
    h}{2}} \\
\ln Y_{j<m}^{(\infty)}&=&
    2\ln \frac{\sin\frac{\pi}{m+2}(j+1)}{\sin\frac{\pi}{m+2}}\label{jlmtt}\\
 \ln Y_{j\geq m}^{(\infty)}&=&2\ln \frac{\sinh\frac{\beta
    h}{2}(j-m+1)}{\sinh\frac{\al \beta h}{2}} \\
\ln \B_n^{(-\infty)}&=& \ln\l[\te^{\frac{\beta h}{2}\,n}\, \frac{\sinh \frac{\beta
    h}{2}(n+1)}{\sinh \frac{\beta h}{2}}\r] \\
\ln\B_n^{(\infty)}&=& \ln\l[\te^{\frac{\al \beta h}{2}\,(n-m)}\, \frac{\sinh \frac{\al\beta
    h}{2}(n-m+1)}{\sinh \frac{\al\beta
    h}{2}}\r]\nn
\ee
\end{subequations}
with $\al=(1-m\g/ \pi)^{-1}$ and analogously for $\BB_n^{(\pm\infty)}=\l.\B_n^{(\pm\infty)}\r|_{h\to-h}$. The limits of $\ln y_j$, $\ln \b_n$, $\ln \bb_n$ are obtained from the above formulas by $y_j=Y_j-1$, $\b_n=\BB_n-1$ and $\bb_n=\BB_n-1$. For $\b_n$, the results take the simple expression 
\be
\ln \b_n^{(-\infty)}&=&\ln\l[\te^{\frac{\beta h}{2}\,(n+1)}\, \frac{\sinh \frac{\beta
    h}{2}n}{\sinh \frac{\beta h}{2}}\r] \nn\\
\ln \b_n^{(\infty)}&=&\ln\l[\te^{\frac{\al \beta h}{2}\,(n-m+1)}\, \frac{\sinh \frac{\al\beta
    h}{2}(n-m)}{\sinh \frac{\al\beta
    h}{2}}\r]\nn
\ee
and similarly for $\bb_n$ with $h\to-h$. 

\begin{figure}[t]
\vspace{-0.6cm}
\includegraphics[angle=-90, width=7.5cm]{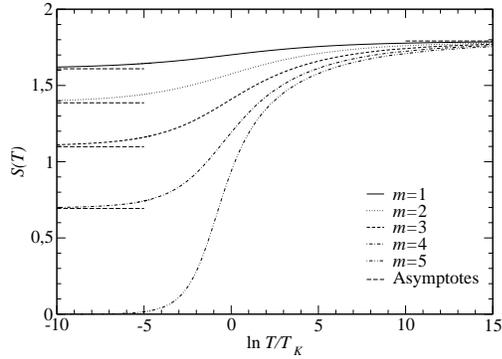}
\vspace{-0.4cm}
\caption{\label{fig1} Entropy for impurity spin $S=5/2$ and different channel numbers. At low temperatures, the asymptotes $\ln(m+1)$ depend on the channel number, at high temperatures, the common asymptote $\ln(2S+1)$ is approached.}
\end{figure}

\begin{figure}[t]
\vspace{-0.6cm}
\includegraphics[angle=-90,width=7.5cm]{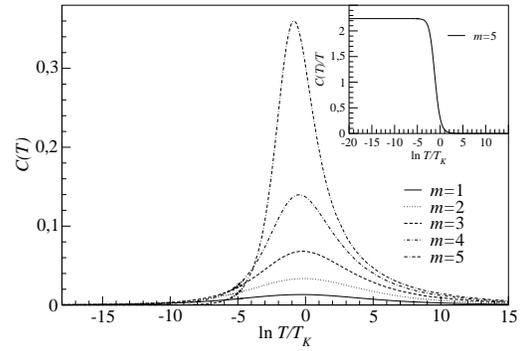}
\vspace{-0.4cm}
\caption{\label{fig2} Specific heat for impurity spin $S=5/2$ and different channel
  numbers. The inset shows the Fermi liquid behavior $\lim_{T\to0}
  C(T)/T=\text{const.}$ for $2S=m=5$. }
\end{figure}

\begin{figure}[t]
\vspace{-0.6cm}
\includegraphics[angle=-90, width=7.5cm]{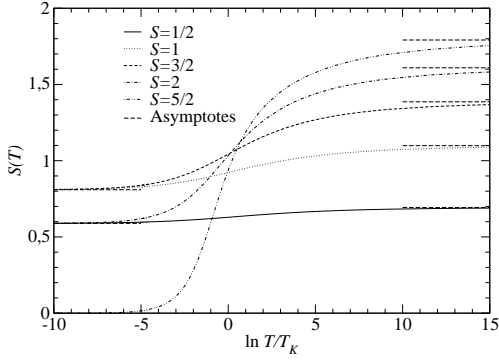}
\vspace{-0.4cm}
\caption{\label{fig3} Entropy for $m=5$ channels and spins $S=1/2,\ldots 5/2$. In the low temperature limit, the entropy takes the unusual value $\ln\l\{\sin\l[\pi(2 S+1)/(m+2)\r]/\sin \pi/(m+2)\r\}$, equation \refeq{jlmtt}, whereas for high temperatures, the asymptotes are given by $\ln (2S+1)$.  }
\end{figure}

\begin{figure}[t]
\vspace{-0.6cm}
\includegraphics[angle=-90, width=7.5cm]{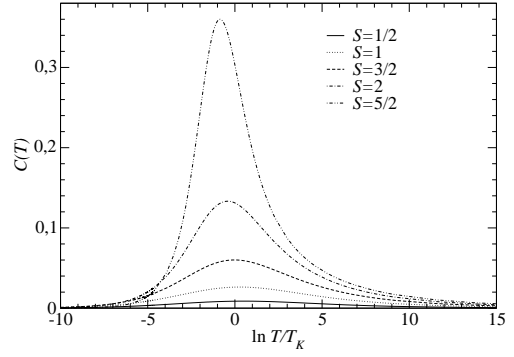}
\vspace{-0.4cm}
\caption{\label{fig4} Specific heat for $m=5$ channels and spins $S=1/2,\ldots 5/2$. }
\end{figure}

\begin{figure}[t]
\vspace{-0.6cm}
\includegraphics*[angle=-90, width=7.5cm]{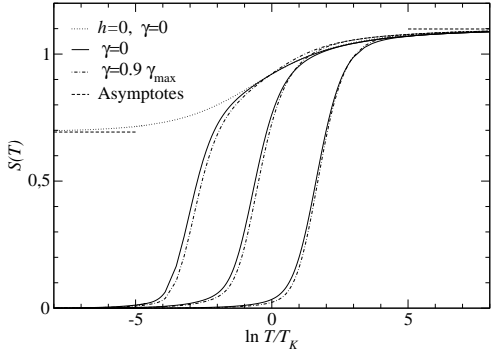}
\vspace{-0.4cm}
\caption{\label{fig5}  The entropy for the under-screened case $S=1,\,m=1$ with fields $h=0,
  0.1,\,1,\,10$ and anisotropies $\g=0,\,0.9 \g_{\max}=0.707$ (for $\g_{\max}$, see equation \refeq{restan}). 
}
\end{figure}

\begin{figure}
\vspace{-0.6cm}
\includegraphics[angle=-90, width=7.5cm]{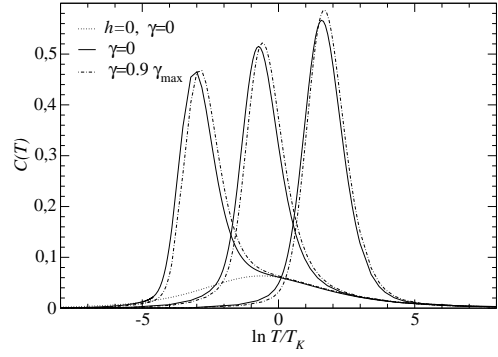}
\vspace{-0.4cm}
\caption{\label{fig6} The specific heat for the under-screened case $S=1,\,m=1$ with fields $h=0,
  \,0.1,\,1,\,10$ for anisotropies $\g=0,\,0.9 \g_{\max}=0.707$ (for $\g_{\max}$, see equation \refeq{restan}).}
\end{figure}

The above presented derivation of the closed set of finitely many integral equations for the impurity's free energy is speculative in the sense that an observation made for the well-understood case $S=1/2$, $m=1$ within a path integral approach \cite{bo04} is extended to the case of general $S$ and $m$. We intend to report on the generalization of  \cite{bo04} in a future publication. Our claim, however, that \refeq{nlis}-\refeq{restdef}, \refeq{mchsspev} hold also in the general case follows by a simpler reasoning. By use of the algebra \cite{jue97eq, suz98} of fused transfer matrices one can show that our key equations are equivalent to the
infinitely many TBA equations of \cite{tsv85} in the isotropic case and to the equations of \cite{sch00,sch01} in the anisotropic case. As an illustration, figures \ref{fig1}-\ref{fig6} show
the entropy and specific heat for some special cases. The curves for $\g=0$
agree nicely with similar graphs in \cite{des85,sch93}; those for $\g\neq 0$ are novel. 

We expect that in the framework of a generalized lattice path integral approach, additional electron-electron interactions in the host must be introduced to keep the model integrable. These interactions may lead to both a renormalization of the Fermi velocity $v_f$ and the electronic $g$-factor, analogous to the known exact solution \cite{tsv83}. These additional interactions will be the concern of future work; the analysis in this work is based on the model \refeq{defhhgen}-\refeq{defhexgen}. 

\section{Low temperature evaluation}
\subsection{$2S=m$: Exact screening}
\label{loword}

The first non-vanishing order in a low temperature, low field expansion of the
free energy in the exactly screened case can be obtained by expanding the corresponding integral in the following form
\be
f_i(T,h)&=&-\frac{T}{2\pi} \i \frac{\ln
\l[\B_l\BB_l\r](x)}{\cosh\l(x+\ln \frac{T}{T_K}\r)}\d x\nn\\
&\stackrel{T,h\ll T_K}{=}&-\frac{T^2}{\pi T_K} \i \te^x \ln
\l[\B_l\BB_l\r](x)\,\d x \label{lef}\; .
\ee
The temperature and the field are supposed to be small compared to $T_K$. As can be seen from equations \refeq{asympval}, the integral in equation \refeq{lef} only exists for $l=m$. The above approximation makes sense if $T,h\ll T_K$. The integral in equation \refeq{lef}
can be done exactly, by a generalization of a method used in
\cite{kl91b} leading to dilogarithms and applied for instance to the spin-$S$ Heisenberg chain in \cite{bab86,suz98}. 

In the notation of equation \refeq{nlis}, consider the integral
\be
I(T,h)\!\!:= \!\!\i \l(\frac{\d}{\d x} {\bf y}(x)\, {\bf Y}(x)-{\bf
  y}(x)\frac{\d}{\d x}{\bf Y}(x)\r) \d x\label{dili}\,.
\ee
From the symmetry of the integration kernels one deduces 
\be
f_i(T,h)=\frac{T^2}{2\pi T_K} \,I(T,h)-\frac{f_\g}{2\pi T_K}\,h^2 m\label{fht}\; .
\ee
On the other hand, the integral $I(T,h)$ is calculated as in \cite{suz98} by using dilogarithmic
identities, the result is
\be
I&=& -\pi^2\,\frac{m}{m+2}\label{res2}\; .
\ee
Combining equations \refeq{fht}, \refeq{res2}, one gets the first term in an
expansion of the free energy for low fields and temperatures:
\be
f_i(T,h)&=&-\frac{T^2}{2T_K}\,\frac{\pi m}{m+2} -
\frac{m}{4(\pi-m\g)}\,\frac{h^2}{T_K}\label{lowft}\; .
\ee
From equation \refeq{lowft},
\be
C_i(T)&=&\frac{T}{T_K}\,\frac{\pi m}{m+2}\qquad \chi_i(h)=\frac{1}{T_K}\,\frac{m}{2(\pi-m\g)}\label{chidl} \; .
\ee
These relations exhibit clearly the role of $T_K$ as a ``low temperature scale''. 
According to equation \refeq{chidl}, the impurity contribution to the specific
heat and magnetic susceptibility are Fermi liquid like at $T,h\ll T_K$ in the
exactly screened case $l=m$. This is the regime of ``strong coupling'': The
anti-ferromagnetic spin exchange leads to the formation of a many particle
state between the impurity and the host electrons, which screens the magnetic moment of the impurity. Elementary excitations of this bound state are
Fermi like. Nozi\`eres \cite{noz74,noz80} built up a phenomenological Fermi liquid theory to
describe this regime. 
This Fermi liquid behavior is to be compared with the host. It consists of
spin-1/2-fermions of $m$ non-interacting
channels (or of $m$ flavors), so that the density of states is enhanced by a factor of
$m$. 
\be
C_{h}(T)&=& T\frac{m\pi^2}{3}\rho_h\qquad 
\chi_h(h)=\frac{m\,\rho_h}{4}\label{trueferm}\; .
\ee
The coefficient of the linear $T$-dependence of $C_i$ ($C_{h}$) is denoted by $\delta_i$ ($\delta_{h}$). The low-temperature Wilson ratio $R$ is defined and calculated as
\be
R&:=&\lim_{T\to 0}\frac{\chi_i}{\chi_h} \,
\frac{\delta_h}{\delta_i}=\frac{2(m+2)}{3\l(1-m\frac{\g}{\pi}\r)}\geq 2\label{defwrlt}\; .
\ee
The lower bound $2$ is reached for $m=1$, $\g=0$. The striking feature in
comparing equations \refeq{chidl}, \refeq{trueferm} is that $C_i$ is reduced by a
factor of $3/(2m+4)$ in comparison to $C_h$, if the constant $\rho_h$ is
chosen such that $\chi_i=\chi_h$ for $\g=0$. This may be interpreted by the
localization of the impurity: Contrary to the host electrons, it does not
move, so that the specific heat is reduced.

\subsection{Linearization}
\label{lin}
In this section, the ground-state and the lowest $T$-dependent contribution to
the free energy are calculated by linearizing the NLIE for $h\neq 0$.  
From equations \refeq{asympval} one observes that for $h\neq 0$ and $\beta \to \infty$ in the limit $x\to-\infty$ the auxiliary functions scale with
$\beta h$. Especially,  $\lim_{\beta\gg1}\ln\BB_{l}=0+\O\l(\te^{-\beta h}\r)$, so that one can neglect $\ln \BB_l$ with exponential accuracy.

One introduces the scaling functions $\epsilon_j$,
\be
\begin{array}{ll}
\ln y_j(x)=:\beta h K_m\,\epsilon_j(-x+\ln\beta hK_m)& j=1,\ldots,l-1\\
\ln \b_l(x)=:\beta
h K_m\,\epsilon_l(-x+\ln\beta hK_m)\; ,& \end{array}\nn
\ee
where $K_m$ is defined by $\ln y_m(\ln\beta hK_m)=0$.
The shift in the spectral parameter is performed in
order to deal with functions which have a zero in the origin, 
\be
\e_m(0)=0\label{zerodef}\; .
\ee
Note that therefore 
\be
\e_j(x)\l\{\begin{array}{ll}
  >0\, ,\; x>\ln K_m\, ;& j=m\\
 < 0\,,\;  x<\ln K_m\,;& j=m\\
  >0\;;&\text{otherwise}\end{array}\r.\; ,
\ee
so that one linearizes the logarithms 
\be
\lefteqn{\ln\l[1+\exp\l(\beta h K_m \e_j(x)\r)\r]}\nn\\
\!\!\!&=&\!\!\!
\l\{\begin{array}{ll}
\beta h K_m \e_j\theta(x)+\O(\text{const.}),\; &j< m\\
\beta h K_m \e_m \theta(x)+\frac{\pi^2}{6\beta h K_m |\e_m(0)|}\,\delta(x),\;&j=m\\
\beta h K_m \e_j+\O(\exp(-\text{const.}\beta h)),\; &j> m
\end{array}\r.\; \label{linlog}
\ee
The linearization is done with exponential accuracy for $\e_{j\geq m}$, but
with only algebraic accuracy for $\e_{j<m}$ because of the constant asymptotes
of the latter functions. The first $T$-contribution for
these is calculated below within a different linearization
scheme. Here, the lowest
$T$-contribution for the exactly and under-screened single
channel case $l\geq m=1$ is determined. For $\g=0$, the results of
\cite{tsv83} are confirmed.

It is convenient to define a matrix $\hat A^{-1}$
\be
\hat A^{-1}_{i,j}(k)\!:=\! \l\{\begin{array}{ll}
    1-\F[\k] &\! i=j=n\\
    \delta_{i,j}-\F[s](\delta_{i,j+1}+\delta_{i,j-1})&\! \mbox{otherwise.}
    \end{array}\r. \label{definvak} \;
\ee
By inserting the $\e$-functions into the original set
of NLIE and using equations \refeq{linlog}, \refeq{definvak} one obtains
\be
\l[\hat A_{ij}^{-1}*\e_{j}\r]&=&\delta_{i,m}\l(\tilde d_m-\e_m^-\r)+\delta_{i,l}\frac{g}{K_m}\label{logsys} ,
\ee
where we have defined the driving term
\be
\tilde d_m(x)&:=&-\te^{-x}+\frac{\pi^2}{6\beta h K_m
  |\e_m'(0)|}\,\delta(x),\nn
\ee
and
\be
\e_{j}(x)=\l\{\begin{array}{l}\e^+_{j,0}(x)  \,,\, j<m\\
\e_{j,0}^+(x) + \frac{\pi^2}{6 (\beta h
  K_m)^2|\e_m'(0)|}\,\l(U^+_{j}(x)+\delta(x)\r),\\
\qquad  j=m \\
\e_{j,0}(x) + \frac{\pi^2}{6 (\beta h
  K_m)^2|\e_m'(0)|}\,U_{j}(x) \,,\, j>m
\end{array}\r.\label{defu}
\ee
The terms $\e_{j,0}$ contain the
ground-state, and from the rest, the lowest
$T$-dependent contributions are obtained. We will calculate the latter
explicitly for
$l\geq m=1$.  We made use of the shorthand-notations
\be
\e_j(x)\,\theta(x)=:\e_j^+(x)\; , \qquad \e_j(x)\,\theta(-x)=:\e_j^-(x)\nn
\ee
for functions in direct space. Their Fourier transforms are denoted as
\be
\F\l[\e_j^+\r]&=&\frac{1}{2\pi} \i \e_j^+(x)\te^{\ti k x}\d x=:\e_j^+(k)\nn\; ,
\ee
hence an index $^+$ denotes analyticity in the upper half of the complex
$k$-plane. 
Define a new energy scale
\be
T_h:=\frac{T_K}{K_m}\label{thdef}\; .
\ee
Then the free energy, magnetization and specific heat are given by
\be
f_i(h)\!\!&=&\!\!-\frac{h K_m}{2}\i \frac{\e_l(k)}{\cosh\frac{\pi k}{2}}\te^{-\ti k \ln
  \frac{h}{T_h}}\,\d k\label{fenlin}\\
M_i(h)\!\!&=&\!\!\frac{K_m}{2}\i\frac{1-\ti k}{\cosh{\frac{\pi k}{2}}} \e_{l,0}(k)\te^{-\ti
  k\ln \frac{h}{T_h}}\d k\label{mag}\\
C_i(T,h)\!\!&=& \!\!\frac{\pi^2T}{6hK_m|\e_m'(0)|}\i \frac{U_l (k)\te^{-\ti k \ln
  \frac{h}{T_h}}}{\cosh\frac{\pi k}{2}}\,\d k \label{sphlowt} .\;\;
\ee
Obviously, the Wilson ratio $C_i/(T\,\chi_i)$ is a function of $h$
alone.  In the following, the magnetization
and the susceptibility are calculated from $\e_{l,0}$, and the
specific heat from $U_l$, defined in equation \refeq{defu}.

\subsection{Under-screened and exactly screened cases, $2S\geq m$}  
The linear system \refeq{logsys} is solved by inverting the matrix $\hat A^{-1}(k)$. This gives the matrix $\hat A(k)$ with
elements
\be
\l[\hat A(k)\r]_{i,j}=\frac{1}{\mbox{det} \hat A^{-1}(k)}\,\alpha_{i,j}(k)\label{defak}\; ,
\ee
where $\alpha_{i,j}$ is the adjunct to $\l[\hat A^{-1}\r]_{i,j}$ in det$\hat
A^{-1}$. Both quantities are seen to satisfy recursion relations, which can
be solved explicitly with the results 
\be
\mbox{det} \hat A^{-1}(k)&=& \frac{\sinh\frac{\pi}{\g} \, \frac{\pi
    k}{2}}{\l(2\cosh\frac{\pi k}{2}\r)^l \sinh \l(\frac{\pi}{\g}-l\r)\frac{\pi
    k}{2}}\label{dlres} \\
\hat A_{j,j}(k)&=& \frac{2 \cosh \frac{\pi k}{2}\,\sinh j \frac{\pi
    k}{2}\,\sinh \l(\frac{\pi}{\g}-j\r)\frac{\pi k}{2}}{\sinh \frac{\pi
    k}{2}\,\sinh\frac{\pi}{\g}\,\frac{\pi k}{2}} \nn\\
\hat A_{j,l}(k)&=& \frac{2\cosh\frac{\pi k}{2}\,\sinh j \frac{\pi
    k}{2}\,\sinh\l(\frac{\pi}{\g}-l\r) \frac{\pi k}{2}}{\sinh \frac{\pi
    k}{2}\, \sinh \frac{\pi}{\g}\, \frac{\pi k}{2} }\nn
\ee

Other matrix elements are not needed. Then 
\be
\frac{\e_m(k)}{\hat A_{m,m}(k)}&=&-\e_m^-(k)+\tilde d_m(k)+\frac{\hat
  A_{m,l}(k)}{\hat A_{m,m}(k)} \,\frac{f_\g}{K_m}\,\delta(k)\label{epsm1}\qquad\\
\e_l(k)&=&\frac{\hat A_{m,l}(k)}{\hat A_{m,m}(k)}\,\e_m(k)+\nn\\
& &\l(\hat
A_{l,l}(k)-\frac{\hat A^2_{m,l}(k)}{\hat A_{m,m}(k)}\r)
\,\frac{f_\g}{K_m}\delta(k)\label{epsl1}
\ee
Eq. \refeq{epsm1} is solved by factorizing $\hat A_{m,m}$ into
two functions $G_{\pm}$, $G_+$ ($G_-$) being analytic in the upper (lower)
half of the complex $k$-plane (Wiener-Hopf factorization),
\be
\hat A_{m,m}(k)&=&G_+(k)\,G_-(k)\label{faceq}\; .
\ee
The result is
\be
\e^+_{m,0}(k)&=&\frac{1}{2\pi}\,\frac{G_+(k)\,G_-(-\ti)}{(k+\ti
  0^+)(k+\ti)}\label{epsm0}\; \\
\l.U_{1}^+(k)\r|_{m=1}+1&=&
\frac{1}{2\pi}\,G_+(k)\,,\, l>1\,,
\label{u1def}
\ee
where $\tilde k$ is the Fourier transform of the kernel defined in
equation \refeq{intkdef} for $l=1$ and 
\be
G_+(k)\!\!&=&\!\! \l(\frac{2\pi m\l(1-\frac{\g}{\pi}
  l\r)}{1-\frac{\g}{\pi}(l-m)}\r)^{\frac{1}{2}} \nn\\
\!\!& &\!\!\times \frac{
  \G\l(1-\ti \frac{k}{2}\r)\G\l(1-\ti\frac{k}{2}\,\frac{\pi}{\g}\r)\te^{-\ti a k}}{\G\l(\frac{1}{2} -\ti \frac{k}{2}\r)\G\l(1-\ti m\frac{k}{2} \r)\,
  \G\l(1-\ti\frac{k}{2}\l(\frac{\pi}{\g}-m\r)\r)}\nn\\
G_-(k)\!\!&=&\!\! G_+(-k)\; .\nn
\ee
The constant $a$ is chosen to be $a=-\frac{\pi }{2 \g} \ln (1-\g m/\pi)+\frac{m}{2} \ln(1-\pi/(\g m))$.
From equation \refeq{epsm1} the constant $K_m$ is found to be
\be
K_m&=& f_\g\, \frac{\hat A_{m,l}(0)}{\hat A_{m,m}(0)}\,\frac{G_-(0)}{G_-(-\ti)}\label{defkm}\;
.
\ee
With this information and equations \refeq{tkdef}, \refeq{thdef} we know the
magnetic field scale $T_h$ whose role is revealed in various expansions of the
magnetization $m(h)$ below. 
%

\subsubsection*{Ground state}
The ground state contribution to the free energy 
equation \refeq{fenlin} is given
by inserting $\e_{l,0}$ from equation \refeq{epsl1} with equation \refeq{epsm0}. 

The magnetization is written down from equation \refeq{mag},
\begin{widetext}
\be
M_i(h)&=&\frac{l-m}{2}\frac{1}{1-m\frac{\g}{\pi}}\nn\\
& &+ \frac{m}{4\pi^{3/2}}\frac{1-\frac{\gamma}{\pi}l}{1-\frac{\g}{\pi}m}\i\ti
\frac{\G\l(\frac{1}{2}+\ti \frac{k}{2}\r)
  \,\G\l(1-\ti\frac{k}{2}\r)\,\G\l(1-\ti\frac{\pi
    k}{2\g}\r)\,\G\l(1+\ti\frac{k}{2}\l(\frac\pi\g-m\r)\r)}{\G\l(1+\ti\frac{k}{2}\l(\frac{\pi}{\g}-l\r)\r)\,\G\l(1-\ti\frac{k}{2}\l(\frac{\pi}{\g}-l\r)\r)\,\G\l(1-\ti\frac{m k}{2}\r)} \frac{\te^{-\ti k \ln
  \frac{h}{T_h}-\ti a k}}{k+\ti 0^+}\d k\label{mag2}\; .
\ee
\end{widetext}
The integral can be calculated by closing the contour in the lower ($h>T_h$)
or in the upper ($h<T_h$) half plane. This results in two power series in
$T_h/h$ for $h>T_h$ and in $h/T_h$ for $h<T_h$ with integer and non-integer powers, depending on the
poles of the integrand. These calculations are straightforward but lengthy. Especially, in the limits $h\ll
T_h$, $h\gg T_h$, one extracts the constant values
\be
M_i(h)&\to&\l\{\begin{array}{l}
  \frac{l-m}{2}\frac{1}{1-m\frac{\g}{\pi}}+\O\l(\l(\frac{h}{T_h}\r)^{2\g/(\pi-m\g)}\r)
  \,;\\
\qquad l>m\;, h\ll T_h\\
  \O\l(\frac{h}{T_h}\r) \,;\,l=m\,, h\ll T_h\\
  \frac{l}{2}+\O\l(\l(\frac{h}{T_h}\r)^{-2\g/\pi}\r)\,; h\gg T_h\;.
  \end{array}\r.\label{man}
\ee
Thus for $h=0$, a non-integer rest-spin remains in the case $\g\neq 0$,
$l>m$ and $h$-dependent corrections are of non-integer powers of $h$. The
constant terms in \refeq{man} agree precisely with the asymptotes, equations \refeq{asympval} and with \cite{sch00,sch01}, where the same model was investigated by TBA-techniques in the limits indicated in \refeq{man}.

In the isotropic limit $\g\to 0$
\be
G_+(k)&=&  \frac{\sqrt{2\pi m} \,\G\l(1-\ti \frac{k}{2}\r)}{\G\l(\frac{1}{2} -\ti
  \frac{k}{2}\r)\,\G\l(1-\ti \frac{k m}{2}\r)}
\l(\frac{-\ti k m}{2\te}\r)^{-\ti\frac{m k}{2}}\nn\\
M_i(h)&=&\frac{m}{4\pi^{3/2}}\i\ti
\frac{\G\l(\frac{1}{2}+\ti \frac{k}{2}\r)
  \,\G\l(1-\ti\frac{k}{2}\r)}{\G\l(1-\ti\frac{m k}{2}\r)}\nn\\
& & \times \l(-\frac{\ti km}{2\te}\r)^{-\ti\frac{km}{2}}\te^{-\frac{\pi|k|}{2}(l-m)}\frac{\te^{-\ti k \ln
  \frac{h}{T_h}}}{k+\ti 0^+}\d k\nn\\
& &+\frac{l-m}{2}\to\l\{\begin{array}{cc}
  \frac{l-m}{2}, \;& h\ll T_h\\
  \frac{l}{2},\;& h\gg T_h
  \end{array}\r.\label{mhlgm} .
\ee
In the last line, only the leading behavior due to the simple pole at $-\ti
0^+$ for high fields has been included. The integral in equation \refeq{mhlgm} has
been given by Tsvelick and Wiegmann, \cite{tsv84}. It allows to determine the
zero-temperature scales defined below. There are several singularities of the integrand equation \refeq{mhlgm} in the complex $k$-plane:
\begin{itemize}
\item[i)] $m=l$: Poles are distributed in the upper and lower half-planes, with an additional dominating cut along the negative imaginary axis.
\item[ii)] $m<l$: A cut along the whole imaginary axis goes along
  with sub-leading poles in both the upper and lower half planes.
\end{itemize}
Singularities in the lower (upper) half plane are relevant for $h>T_h$
($h<T_h$). Poles $k_n=\ti (2n+1),\; n=0,1,\ldots$ in the upper half plane only give a leading contribution in the
exactly screened case $l=m$. They have residuals $2\ti(-1)^{n+1}/n!$, such that the
magnetization is given by a series
\be
M_i(h<T_h)&\stackrel{m=l}{=}& \frac{m}{\sqrt
  \pi}\sum_{n=0}^\infty \frac{(-1)^n}{(2n+1)n!}\frac{\G\l(n+\frac12\r)}{\G\l(m\l(n+\frac12\r)\r)}\nn\\
& &\times \l(\frac{m\l(n+\frac12\r)}{\te}\r)^{m\l(n+\frac12\r)}\l(\frac{h}{T_h}\r)^{2n+1}\nn\; .
\ee
One recognizes the signature of a Fermi liquid in first order : $M_i(h)\propto h$ and
$\chi_i(h=0)=$const. Upon inserting explicit values, one finds agreement with
equation \refeq{chidl}. We shall establish this agreement explicitly for arbitrary $\g$ below, equations \refeq{mlm}, \refeq{chilm}.

Let us draw our attention to the cut in the lower half plane, for values
$l\geq m$. By linearizing the integrand, we find
\be
M_i(h>T_h)-\frac l2\!\!\!&\stackrel{l\geq m}{=}&\!\!\! -\frac{ml}{4}\l(\frac{1}{\ln
  \frac{h}{T_h}}+\frac{m}{2}\frac{\ln\ln\frac{h}{T_h}}{\ln^2
  \frac{h}{T_h}}\r.\nn\\
\!\!\!& & \!\!\!\l.+\frac{A}{\ln^2\frac{h}{T_h}}+\O\l(\ln^{-3}h/T_h\r)\r)\nn\\
A\!\!\!&=&\!\!\!\frac 12\l[-m\ln m+(m-2)\ln2\r]\label{defat0}\, .
\ee
The contribution $\O(\ln^{-2}h/T_h)$ is absorbed by the definition 
\be
T_h&=& \te^{-A} \widetilde T_h\label{deftildeht0}\\
M_i(h>\widetilde T_h)-\frac l2&\stackrel{l\geq m}{=}& -\frac{ml}{4}\l(\frac{1}{\ln
  \frac{h}{\widetilde T_h}}+\frac{m}{2}\frac{\ln\ln\frac{h}{\widetilde T_h}}{\ln^2
  \frac{h}{\widetilde T_h}}\r.\nn\\
& & \l.+\O\l(\ln^{-3}h/\widetilde T_h\r)\r)\label{mlgeqmunih}
\ee
Equation \refeq{deftildeht0} defines the zero-temperature scale for large magnetic
fields. 

Finally, upon ''encircling'' the cut in the upper half plane, one only
replaces the pre-factor $l$ in equation \refeq{mlgeqmunih} by $l-m$ and sets $-\ln
h/\widetilde T_h=\ln \widetilde T_h/h$:
\be
M_i(h<\widetilde T_h)\!\!\!&\stackrel{l>m}{=}&\!\!\!\frac{l-m}{2}+ \frac{m(l-m)}{4}\l(\frac{1}{\ln
  \frac{\widetilde T_h}{h}}\r.\nn\\
\!\!\!& &\!\!\! \l.-\frac{m}{2}\frac{\ln\ln\frac{\widetilde T_h}{h}}{\ln^2
  \frac{\widetilde T_h}{h}}+\O\l(\ln^{-3}\widetilde T_h/h\r)\r)\label{mlgeqmunil}
\ee 
Note that this simple replacement does not hold for $\g\neq 0$, as can already be seen from the lowest
order, equation \refeq{man}. 

The free spin value of the magnetization is approached logarithmically at high
fields. This asymptotical freedom in the weak coupling limit is a
genuine feature of the Kondo model. An analogous effect occurs for low
fields in the under-screened case; however, the first correction is of
opposite sign compared to the high-temperature case,
cf. equations \refeq{mlgeqmunih}, \refeq{mlgeqmunil}. Classical Fermi liquid behavior
appears at low temperatures if the impurity is exactly screened. The physical origin of these results has been revealed by Nozi\`eres \cite{noz80} already before the exact solution
of the Kondo model was known: At high fields, corrections to the asymptotical
freedom of the impurity spin are caused by the weak anti-ferromagnetic coupling with
the host particles. At low fields, the impurity spin is partially screened due
to strong anti-ferromagnetic exchange. Two kinds of interactions with this
impurity-electron system may occur: On the one hand, a weak {\em ferro}magnetic coupling of the
residual spin $S-m/2$ with the host, due to the Pauli principle (this explains
the change of sign in the leading order on the rhs of equations \refeq{mlgeqmunih}, \refeq{mlgeqmunil}). On the other
hand, a polarization of the bound complex by host electrons, analogously to
the Fermi liquid excitations at $S=m/2$. This polarization is given by the poles and
dominated by the ferromagnetic Kondo interactions, reflected by the cuts in
the complex plane.  
\subsubsection*{Finite temperature}
After the calculation of the magnetization, we proceed with the specific heat
for $l\geq m=1$. By inserting equation \refeq{u1def} into equation \refeq{sphlowt},
we get 
\be
C_i(T\ll T_K,h)/T=\frac{\pi^2}{3hK_1|\e_1'(0)|}V(-\ln h/T_h)\nn\;,
\ee
where the Fourier transform of $V(x)$ is given by
\be
V(k)=\l\{\begin{array}{ll} s(k)\l[U_1^+(k)+1\r]\,,& l=1\\
 \frac{A_{1l}(k)}{A_{11}(k)}s(k)\l[U_1^+(k)+1\r] \,,&l\geq 1
\end{array}\r.\nn\; .
\ee
From equation \refeq{u1def} it follows for $l=1$ that
$V(k)= G_+(k) s(k)$, so that
\be
C_i(T\ll T_K,h)/T=\frac{\pi^2}{3|\e_1'(0)|G_-(-\ti) K_1^2}\chi(h)\label{cingfun}\;,
\ee
where we have inserted equations \refeq{mag}, \refeq{epsm0}, \refeq{u1def}. The constant
$|\e_1'(0)|$ is determined from the condition that equation \refeq{cingfun}
matches equation \refeq{chidl} for $l=1=m$, so that $|\e_1'(0)|=1/(K_1G_-(0))$, and
\be
C_i(T\ll T_K,h)/T=\frac{2\pi(\pi-l \g)}{3}\chi(h)\label{sphchirel}\;.
\ee

\subsection{Over-screened and exactly screened cases, $2S\leq m$ (ground state)}
\label{seclleqm}
In this section, we calculate the ground state for $l\leq m$, finite
temperatures are dealt with in the next section. 

If $l\leq m$, there are $m$ equations, the last one being
\be
\e_m(x)&=&-\te^{-x} +\frac{f_\g}{K_m}+[s*\e_{m-1}+\k*\e_m](x)\nn\; ,
\ee
where $f_\g=\frac{1}{2(1-m\g/\pi)}$. On the right-hand side of equation \refeq{logsys}, only the $m^{th}$ entry
is different from zero. Consequently,
\be
\frac{\e_m^+(k)}{\hat A_{m,m}(k)}&=& -\e_m^-(k)+\tilde d_m(k)+\frac{f_\g}{K_m}\delta(k)\label{epsm2}\\
\e_l&=& \l[\frac{\hat A_{l,m}}{\hat A_{m,m}}\r]\,\e_m^+ \label{elp1}
\ee
Equations \refeq{epsm1}, \refeq{epsm2} are solved by
the Wiener-Hopf method, afterwards $\e_m^+$ is inserted in equations \refeq{epsl1},
\refeq{elp1}. The relevant matrix entries read
\be
\hat A_{m,m}(k)&=&\frac{2\cosh\frac{\pi}{2} k\, \sinh{ m\frac{\pi k}{2}} \, \sinh
  \frac{\pi k}{2}\l(\frac\pi\g -m\r)}{\sinh\frac{\pi k}{2}\, \sinh\frac\pi\g\frac{\pi
    k}{2}}\nn\\
\hat A_{l,m}(k)&=&\frac{2\cosh\frac{\pi}{2} k\, \sinh{ l\frac{\pi k}{2}} \, \sinh
  \frac{\pi k}{2}\l(\frac\pi\g -m\r)}{\sinh\frac{\pi k}{2}\, \sinh\frac{\pi
    k}{2}\frac\pi\g}\nn\; .
\ee
With $\hat A_{m,m}(k)=G_+(k)\,G_-(k)$, the auxiliary functions are given by
\be
\e_m^+(k)&=&\frac{1}{2\pi}\,\frac{G_+(k)\,G_-(-\ti)}{ (k+\ti 0^+)(k+\ti)}\nn\\
G_+(k)&=&\frac{\l(2m\pi\l(1-m\frac{\g}{\pi}\r)\r)^{\frac{1}{2}}\G\l(1-\ti\frac{k}{2}\r)}
  {\G\l(\frac{1}{2}-\ti
  \frac{k}{2}\r)\,\G\l(1-\ti
  m\frac{k}{2}\r)}\nn\\
& &\times\frac{\G\l(1-\ti\frac{\pi}{\g}\,\frac{k}{2}\r)\te^{-\ti a k}}{\G\l(1-\ti\frac{k}{2}\l(\frac{\pi}{\g}-m\r)\r)}\nn,
\ee
and $K_m=f_\g\,\frac{G_-(0)}{G_-(-\ti)}$.
In the limit of high fields $h\gg T_h$, the
magnetization reads
\begin{widetext}
\be
M_i(h)\!\!\!\!&=&\!\!\!\! \frac{l}{4\pi^{3/2}}\,\i \ti\frac{\G\l(\frac{1}{2}+\ti
  \frac{k}{2}\r)\,\G\l(1+\ti m\frac{k}{2}\r)\,\G\l(1-\ti
  \frac{k}{2}\r)\,\G\l(1-\ti \frac{k}{2}\,\frac{\pi}{\g}\r)}{\G\l(1-\ti
  \frac{k}{2}\,\l(\frac{\pi}{\g}-m\r)\r)\,\G\l(1-\ti
  l\frac{k}{2}\r)\,\G\l(1+\ti l\frac{k}{2}\r)}\frac{\te^{-\ti k \ln
  \frac{h}{T_h} -\ti a k}}{k+\ti 0^+}\d k\stackrel{h\gg T_h}{\to} \frac{l}{2}\label{mhllm}\; .
\ee
\end{widetext}
The isotropic limit $\g \to 0$ yields 
\be
G_+(k)&=&\frac{\sqrt{2m\pi}\,\G\l(1-\ti\frac{k}{2}\r)}{\G\l(\frac{1}{2}-\ti\frac k2\r)\,\G\l(1-\ti m \frac k2\r)}\l(-\frac{\ti km}{2\te}\r)^{-\ti \frac{m k}{2}}\nn\\
M_i(h)&=&\frac{l}{4\pi^{3/2}}\i \ti\frac{\G\l(\frac{1}{2}+\ti
  \frac{k}{2}\r)\,\G\l(1+\ti m\frac{k}{2}\r)\,\G\l(1-\ti
  \frac{k}{2}\r)}{\G\l(1-\ti
  l\frac{k}{2}\r)\G\l(1+\ti l\frac{k}{2}\r)}\nn\\
& & \times \l(-\frac{\ti km}{2\te}\r)^{-\ti \frac{m k}{2}}\,\frac{\te^{-\ti k \ln
  \frac{h}{T_h}}}{k+\ti 0^+}\d k\nn
\ee
Note that for $\g=0$, a cut along the negative part of the imaginary axis
occurs. It dominates the poles in the lower half of the complex plane, and one is faced with the expected Kondo behavior for $h>T_h$, again in accordance with \cite{tsv84}. 

For $h<T_h$, the leading behavior is given by the
poles with smallest positive imaginary part. 
\be
M_i(h) \propto \l(\frac{h}{T_K}\r)^{2/m},\chi_i(h) \propto  \l(\frac{h}{T_K}\r)^{2/m-1}\label{chiovt0}\; .
\ee
Thus over-screening induces non-integer exponents of $h$, independent of $\g$, at $T=0$. 

Finally note that for $l=m$, the two expressions equations \refeq{mhlgm},
\refeq{mhllm} coincide. It is shown that the first non-vanishing order linear
in $h$ of the magnetization leads to the $T=0$, $h=0$ susceptibility,
calculated in equation \refeq{chidl}. For $l=m$, the magnetization reads
\be
M_i(h)&=& \frac{m}{4\pi^{3/2}}\i\ti
\frac{\G\l(\frac{1}{2}+\ti \frac{k}{2}\r)
  \,\G\l(1-\ti\frac{k}{2}\r)\,\G\l(1-\ti\frac{\pi k}{2\g}\r)}{\G\l(1-\ti\frac{m k}{2}\r)\,\G\l(1-\ti\frac{ k}{2}\l(\frac{\pi}{\g}-m\r)\r)}\nn\\
& &\times \frac{\te^{-\ti k \ln
  \frac{h}{T_h}-\ti ak}}{k+\ti 0^+}\d k\nn\; .
\ee
Being interested in fields $h<T_h$, one takes account of the poles at
$k_n:=\ti(2n+1)$ with residuals $2\ti (-1)^{n+1}/n!$. These result in the series
\be
\lefteqn{M_i(h)= \frac{1}{\pi^{1/2}\l(1-\frac{\g}{\pi}m\r)}\,\sum_{n=0}^{\infty}
\frac{(-1)^n}{(2n+1)n!}}\nn\\
& &\times
\frac{\G\l(\frac{1}{2}+n\r)\G\l(\frac{\pi}{\g}\l(n+\frac{1}{2}\r)\r)a^{2n+1}}{\G\l(m\l(\frac{1}{2}+n\r)\r)\G\l(\l(\frac{\pi}{\g}-m\r)\l(n+\frac{1}{2}\r)\r)}\,\l(\frac{h}{T_h}\r)^{2n+1}\nn\;
.
\ee
From the definition of $T_h$, equation \refeq{thdef}, one gets to first order in
$h$:
\be
\lim_{h\ll T_h} M_i(h)&=& \frac{m}{2\pi\l(1-\frac{\g}{\pi}m\r)}\,\frac{h}{T_K}\label{mlm}\\
\chi_i(T=0, h=0)&=& \frac{m}{2\pi\l(1-\frac{\g}{\pi}m\r)}\,\frac{1}{T_K}\label{chilm}\; .
\ee
This result is expected from equation \refeq{chidl}.
\subsubsection*{Finite temperature}
\label{asunder}
As pointed out above, we are not able to account for
corrections of the linearized functions $\e_{j<m}$, $m>1$ in the framework
of the rigorous linearization. In the following, we consider the case $j<m$ for small
magnetic fields and low temperatures. For $\g=0$, this case was treated previously by Affleck
\cite{aff91} using bosonization and CFT-techniques. He calculated the
low-temperature Wilson-ratio analytically for arbitrary $m$. Analogous results
have been obtained by the TBA-solution\cite{sac89} only for $m=2$, $S=1/2$. Here, we will confirm Affleck's findings for $m\geq 2$, extended to the anisotropic case $\g\neq 0$. 

First consider the case $m=2,\, 2S=1$. The relevant auxiliary functions are
given by:
\be
\ln y_1(x)&=&\l[s*\ln \B_2 \BB_2\r](x)\nn\\
\ln \b_2(x)&=&-\te^x+\frac{\beta h}{2} + [s*\ln Y_1](x)+\l[\k*\ln \B_2\r](x)\nn\\
& &-\l[\k_-*\ln
\BB_2\r](x)\nn\\
f_i(T,h)&=&-\frac{T}{2\pi}\,\i\frac{\ln Y_1(x)}{\cosh\l(x+\ln
  \frac{T}{T_K}\r)}\, \d x\label{fem2}\; .
\ee
In section \ref{loword}, we found (equation \refeq{lowft} specialized to $m=2$):
\be
\lim_{x\gg0} [s*\ln \B_2\BB_2](x)&=& \te^{-x}\l(\frac\pi4-\frac{(\beta
  h)^2}{2(\pi-2\g)}\r)\nn\; .
\ee
Consequently, 
\be
\lim_{x\gg0}\ln Y_1(x)&=& \ln 2 +\te^{-x}\l(\frac\pi8-\frac{(\beta
  h)^2}{4(\pi-2\g)}\r)\label{lnyap}\;
\ee
One approximates the kernel in the same scheme as in
equation \refeq{lowft},
\be
\lim_{T\ll T_K}\frac{1}{\cosh\l(x+\ln T/T_K\r)}&=&
  \frac{2T}{T_K}\,\te^x , \, x\leq \ln \frac{T}{T_K}\label{lncosap} .
\ee
The approximations \refeq{lnyap}, \refeq{lncosap} lead to a constant
integrand, which we deal with by introducing a cutoff $D$, 
\be
 \lefteqn{f_i(T,h)=-\frac{T}{4\pi}\ln 2}\nn\\
& & -\frac{T^2}{T_K}\l[\frac{1}{8}+\frac{(\beta h)^2}{4\pi(\pi-2\g)
  }\r]\int_D^{\ln T/T_K}1 \d x\;
,\; D\ll 0\nn
\ee
The cutoff $D$ depends neither on $T$ nor on $h$, so the
occurrence of the logarithm $\ln T/T_K$ in the above equation leads to 
\begin{subequations}
\be
\lim_{T,h\ll T_K}C_i(T,h)&=&-\frac{1}{4} \frac{T\ln T/T_K}{T_K}\label{cm2}\\
\lim_{T,h\ll T_K}\chi_i(T,h)&=&-\frac{1}{2\pi(\pi-2\g)}\frac{\ln
  T/T_K}{T_K}\label{chim2}\\
\l.R_w\r|_{m=2,2S=1}&=&\frac{8}{3(1-2\g/\pi)}\label{formfl22}\; ,
\ee
\end{subequations}
confirming, for $\g=0$, numerical findings by Sacramento \cite{sac89} and
Affleck's prediction.

For values $m>2$, we perform an asymptotic linearization of the
NLIE in the region $x\to\infty$, following \cite{tsv84}. Corrections to $\ln Y_j^{(\infty)}$ are expressed through a correction function $D_j(x)$, 
\be
\ln Y_j(x)= \ln Y_j^{(\infty)}+D_j(x)\; , \lim_{x\to\infty}D_j(x)=0\nn\; .
\ee
Linearizing $\ln y_j$ to first order in $D_j$, $j=1,\ldots,m-1$
\be
\ln y_j(x)&=&\frac{1}{2} \,\ln Y_{j-1}^{(\infty)}Y_{j+1}^{(\infty)} + \frac{f_j^2}{f_{j-1}\,f_{j+1}}\, D_j(x)\nn\\
f_j&:=& \l(Y_j^{(\infty)}\r)^{1/2}=\frac{\sin\frac{\pi}{m+2}(j+1)}{\sin\frac{\pi}{m+2}}\nn\\
D_j(x)&=&\frac{f_{j-1}\,f_{j+1}}{f_j^2}\l[s*\l(D_{j-1}-D_{j+1}\r)\r](x) \label{aslin}\\
D_m&\equiv& \ln \B_m\BB_m\label{dbbm}\;.
\ee
Equation \refeq{aslin} has been derived for $x\gg 0$. This is
approximately accounted for by writing $D_j(x)\equiv D_j(x)\,\theta(x)$. The
linearized equations form an algebraic system by Fourier transforming. Each unknown function $D_j(x)$ can be expressed in terms of $D_m$,
\be
\!\!\!\!D_j(x)\!\!&=&\!\!\l[t_j*D_m\r](x)\label{tjas}\\
\!\!\!\!t_j(k)\!\!&:=&\!\! \frac{f_{j-1}\sinh(j+2)\frac{\pi k}{2} -f_{j+1}\sinh j\frac{\pi k}{2}}{f_j}\,g(k)\label{tjkdef} .
\ee
As usual, $\F\l[t_j(x)\r]:=t_j(k)$. We see that $D_j$ from equation \refeq{tjas} satisfies equation \refeq{aslin} for $j=1,\ldots,m-2$. The function $g(k)$ has to be determined from the last equation $j=m-1$ and is found to be
\be
g(k)&=& \l[2 \cos\frac{\pi}{m+2} \, \sinh(m+2)\frac{\pi k}{2}\r]^{-1}\label{defgk}\; .
\ee
The first equation, $j=0$, is already contained in equation \refeq{tjas} by
$t_0\equiv 0$ and therefore $D_{j=0}\equiv 0$. Equations \refeq{tjkdef},
\refeq{defgk} imply that the auxiliary functions $\ln Y_j$ behave as
\be
\ln Y_j(x\to\infty)\sim \ln Y_j^{(\infty)} +\text{const.}\times\te^{-4 x/(m+2)}\; .
\ee
The constant is given through $\ln \B_m\BB_m$, equations \refeq{dbbm}, \refeq{tjas}, and will be determined numerically
below. 
Combining equations \refeq{tjas}, \refeq{tjkdef} and \refeq{defgk}, one finds for the impurity part of the free energy
\be
\lefteqn{f_i(T,h\ll T_K)= - T \ln \frac{\sin\frac{\pi(2S+1)}{m+2}}{\sin \frac{\pi}{m+2}}}\nn\\
& &-\frac{T}{2}\i \frac{1}{\cosh\frac{\pi k}{2}} \,D_{2S}(k)\,\te^{\ti k \ln T/T_K}\,\d k\label{fovsc}\; .
\ee
Since $T<T_K$, only negative imaginary values for $k$ are
allowed.

The leading $T$-contribution is given by the singularity closest to the real
axis of the integrand in equation \refeq{fovsc}. For $m\geq 3$, this is a simple
pole at $k_p=-4\ti/(m+2)$, so that
\be
\lefteqn{f_i(T,h\ll T_K)= - T \ln \frac{\sin\frac{\pi(2S+1)}{m+2}}{\sin \frac{\pi}{m+2}}-T\alpha\l(\frac{T}{T_K}\r)^{\frac{4}{m+2}}\hspace{1cm}}\nn\\
\alpha\!&=&\!\!\! \frac{\l(\sin\frac{4\pi S }{m+2}\,\sin\frac{2\pi
    (S+1)}{m+2}-\sin\frac{2\pi S}{m+2}\sin\frac{4\pi
    (S+1)}{m+2}\r)}{(m+2)\cos\frac{\pi}{m+2}\cos\frac{2\pi
  }{m+2}\,\sin\frac{\pi (2S+1)}{m+2}}\nn\\
& & \times\i\te^{\frac{4 x}{m+2}}\,\ln[\B_m\BB_m](x)\,\d x\hspace{1cm}\label{alphap}
\ee
The specific heat and susceptibility are derived:
\begin{subequations}
\be
C_i(T,h=0)&=& \l.\alpha\r|_{h=0} \frac{4(m+6)}{(m+2)^2}\,\l(\frac{T}{T_K}\r)^{\frac{4}{m+2}}\label{ctsv}\\
\chi_i(T,h=0)&=&
\l(\frac{T}{T_K}\r)^{\frac{4}{m+2}-1}\l.\frac{\partial^2}{\partial(\beta
  h)^2}\alpha\r|_{h=0}\label{chitsv}
\ee
\end{subequations}
The derivative in equation
\refeq{chitsv} only acts upon the integrand in equation \refeq{alphap}. Both $C_i/T$
and $\chi_i$ show the same $T$-dependence, yielding the low-temperature Wilson ratio, defined in equation \refeq{defwrlt}:  
\be
R_w=\frac{\chi_i(T)}{C_i(T)}\frac{C_h(T)}{\chi_h(T)}&=&\frac{4\pi^2}{3}\,\frac{(m+2)^2}{4(m+6)}\,\l.\frac{\frac{\partial^2\alpha}{\partial(\beta
    h)^2}}{\alpha}\r|_{h=0}\label{formfl}
\ee
for $m\geq3$. The quantity $\al$ defined in \refeq{alphap} depends on $S,m,\g$ and $\beta h$, which is taken to be zero in \refeq{ctsv}, \refeq{chitsv}. This means that the $\al$-factors in \refeq{ctsv}, \refeq{chitsv} are constant and known, so that from measuring $C_i$ or $\chi_i$, the scale $T_K$ can be determined. So the parameter $T_K$ has the meaning of a low-temperature scale also in the overscreened case. 

\begin{figure}[h]
\vspace{-0.2cm}
\includegraphics*[angle=-90, width=7.5cm]{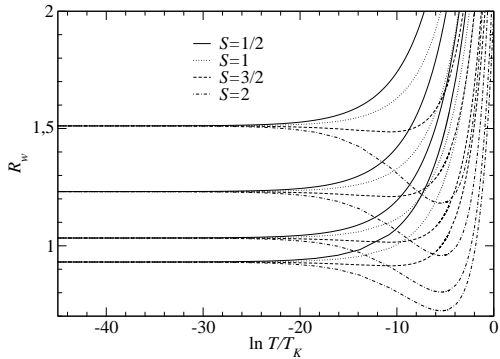}
\vspace{-0.4cm}
\caption{\label{wrovdivanch5} The low-temperature Wilon ratios for the overscreened case with $m=5$
  and anisotropy $\g=0,\,0.2,\, 0.5,\,0.8 \g_{\max}$, where the lowest set of curves is for $\g=0$. }
\end{figure}

We did not find an analytical expression of the integral in
equation \refeq{alphap}, but determined the coefficients in equations \refeq{ctsv},
\refeq{chitsv} numerically. Therefore, the NLIE are solved as described in the
appendix, and the coefficient of the leading $T^{4/(m+2)}$-decay of $C_i(T\ll
T_K)$, $\chi_i(T\ll T_K)\cdot T$ is extracted from the numerical data. In
fig. \ref{wrovdivanch5} some curves are
shown. The Wilson ratio for the isotropic case is 
\be
\l.R_w\r|_{\g=0}&=& \frac{(2+m/2)(2+m)^2}{18}\,;\; m\geq 2\label{affiso}\; .
\ee
For finite anisotropy $\g\neq 0$, the magnetic field is scaled by the factor
$1/(1-m \g/\pi)$ in the NLIE, equation \refeq{intkdef}, \refeq{restdef}. Furthermore, the non-integer powers in \refeq{ctsv},
\refeq{chitsv} are determined by the fusion hierarchy, \refeq{defgk}, where
the anisotropy parameter $\g$ does not enter explicitly. We
therefore conjecture that in analogy to the exactly screened case \refeq{defwrlt} and two-channel overscreened case \refeq{formfl22}, the Wilson ratio for finite $\g$ is   
\be
\l.R_w\r|_{\g}&=&\frac{1}{1-m\g/\pi }\frac{(2+m/2)(2+m)^2}{18}\,;\;
m\geq 2\label{affaniso}\; .
\ee
This means that anisotropy is irrelevant in the renormalization group
sense, in accordance with previous treatments of the underscreened
case by TBA \cite{sch00,sch01,zar02}. On the other hand it has been shown
in the CFT-approach \cite{aff92} that a generic anisotropic spin-exchange is
relevant. Since the spin-exchange \refeq{defhsdgen} is a polynomial in the
spin-operators, the situation here is different from that in \cite{aff92}. We leave
further investigations on this issue as a future task. In table \ref{tabaff}, our results are compared with equations \refeq{affiso}, \refeq{affaniso}.

\begin{table*}
\caption{\label{tabaff} Numerical results for the low temperature Wilson ratios in the over-screened case $m>2S$. In the isotropic case, we compare our data with Affleck's CFT approach \cite{aff91}, equation \refeq{affiso}. For finite anisotropy, the values given in brackets are those from equation \refeq{affaniso}. }
\begin{ruledtabular}
\begin{tabular}{lccccc}
\rule[-3mm]{0mm}{7mm}$m$ & $\g=0$&$\l[1\r]$&$\g=0.2$&$\g=0.5$&$\g=0.8$\\
\hline
\hline
2   &0.202638& 0.202642&  0.2251582 (0.2251582) &0.2071899 (0.2071898)  & 0.3377373 (0.3377373) \\ 
\hline
3   &0.370 & 0.369  &   0.4102 (0.4104) &0.490 (0.493) & 0.607 (0.616) \\
\hline
4   & 0.60791    &0.60793  &  0.6749 (0.6755)   &0.805 (0.811) & 0.99  (1.01)\\
\hline
5   & 0.928 &0.931  &  1.033 (1.034) & 1.23 (1.24) & 1.51 (1.55)    \\
\end{tabular}
\end{ruledtabular}

\end{table*}
%

\section{High Temperature Evaluation}
\label{htgen}
In this section, the NLIE are analyzed in the limit of high temperatures $T\gg
T_K$ by an asymptotic linearization. The fact that contrary to the TBA
approach, we deal with a finite set of NLIE, is most useful here. In the
isotropic single channel case, our results are farther reaching than those obtained from the TBA equations \cite{tsv83}, the treatment of the isotropic multichannel case and of the anisotropic models in this regime is novel. 

This section is divided into three parts. In the first part, the isotropic limit
$\g=0$ is considered. The second part treats the low-temperature limit
of the under-screened case, which is conceptually very similar to the regime of
high temperatures. The anisotropic case is treated in the last part. 

\subsection{Isotropic case $\g=0\,,\;h=0$}
\label{sectgtk}
In the high
temperature regime, the first corrections to the asymptotic value of the free
energy are given by the asymptotic corrections of the concerned
auxiliary functions for $x\to-\infty$. Corrections are of algebraic and logarithmic-algebraic nature and
can be found within an asymptotic linearization
scheme. This is also true for $l>m$ in the low-temperature
limit, referring to the region $x\to\infty$. In this case, the set of $l$
equations decouples into two sets of $m$ and $l-m$ equations and the high-temperature-linearization is
directly transferable.  

One makes use of the knowledge of the asymptotic behavior of the auxiliary
functions derived in appendix \ref{appa} to extract the high-temperature behavior of the free energy, the specific heat and
magnetic susceptibility in zero field. Employing once again the
approximation equation \refeq{crucapp}, this time in the integral of the free
energy, we find
\begin{subequations}
\be
\!\!\lim_{T\gg T_K}f_i(T)&=&
-T\ln(l+1)+\frac{ml(l+2)\pi}{12}\frac{T}{\ln^3\frac{T}{T_K}}\label{ffasht}\\
\!\!\lim_{T\gg T_K} C_i(T)&=&
\frac{1}{\ln^4\frac{T}{T_K}}\,\frac{ml(l+2)\pi^2}{4}\label{casht}\\
\!\!\lim_{T\gg T_K} \chi_i(T)&=&\frac{l(l+2)}{12 T}\l(1-\frac{m}{\ln
  \frac{T}{T_K}}-\frac{m^2 \ln \ln T/T_K}{2\ln^2 \frac{T}{T_K}}\r.\nn\\
& &\l.+\frac{4m^3\phi(m)
    +m^2/4}{\ln^2\frac{T}{T_K}}\r)\label{chiasres}
\ee
\end{subequations}
Values for $\phi(m)$ are given in table \ref{tabbmfin} for $m=1,\ldots,5$. Let us cite the special case $m=1$,
\be
\phi(m=1)=0.04707\pm 2\cdot 10^{-7}\label{numphi}
\ee
Equations \refeq{ffasht}-\refeq{chiasres} constitute the first calculation of the leading orders of the specific heat and magnetic susceptibility for
general $m$ in the framework of an
exact solution. The single-channel case $m=1$ agrees with known TBA-results,
\cite{tsv83}. Especially, the coefficient of the $(\ln
T/T_K)^{-2}$-decay of the magnetic susceptibility is determined, from which we will calculate Wilson's number, relating
the high-temperature to the low-temperature scale.

From equation \refeq{chiasres}, we calculate Wilson's number for the underscreened
and exactly screened cases, $l\geq m$.
Let us first draw our attention to the exactly screened spin-$1/2$ case,
$l=1=m$. Wilson \cite{wil75} obtained for the zero-field spin-$1/2$ susceptibility by his renormalization group approach
\be
\lim_{T\to 0}\chi_i(T)&=&\frac{1}{2\pi T_K} = \frac{0.1032\pm 0.0005}{\widetilde T_K} \nn\\
\frac{\widetilde T_K}{T_K} &=& 2\pi \cdot(0.1032\pm 0.0005)=: 2\pi\xi\label{wrwil}\; .
\ee
This ratio relates the low-temperature scale $T_K$ to the
high-temperature scale $\widetilde T_K$, defined by absorbing the term
$\O(x^{-2})$ in the asymptotic expansion
\be
\lefteqn{\lim_{T\gg T_K}\chi_i(T)= \frac{1}{2\pi T}\,\i\frac{\partial^2_{\beta h}\l[\ln \B_1\BB_1\r](x)}{\cosh\l(x+\ln T/T_K\r)}\d x}\nn \\
&=& \frac{1}{2 T}\partial^2_{\beta h}\l[\ln \B_1\BB_1\r](x=-\ln T/T_K) \nn\\
&=&\l.b_{1,0}^{(\chi)}+ \frac{b_{1,1}^{(\chi)}}{x}+\frac{ b_{1,1}^{(\chi)}\, \ln|x|}{x^2} +\frac{ b_{1,2}^{(\chi)}}{x^2}\r|_{x=-\ln T/T_K} \label{exchi}\\
&=& \l.b_{1,0}^{(\chi)}+ \frac{b_{1,1}^{(\chi)}}{x-\frac{b_{1,2}^{(\chi)}}{b_{1,1}^{(\chi)}}}+\frac{ b_{1,l}^{(\chi)}\, \ln|x|}{x^2}\r|_{x=-\ln T/T_K} \nn\; .
\ee
Wilson's number is identified to be
\be
2\pi \xi=\exp\l(-\frac{b_{1,2}^{(\chi)}}{b_{1,1}^{(\chi)}}\r)\label{wildef}\; .
\ee
Andrei and Lowenstein \cite{and80} carried out a perturbative expansion of the
$2S=m=1$ free energy, both for $ h/T \gg 0$ and $T/h\gg 0$. By requiring that in the
first case, the result should depend on $h/T_h$, in the second case on
$T/\widetilde T_K$, they deduced the ratio $T_h/\widetilde T_K$. Moreover, they
determined the ratio
$T_h/T_K$ from the (conventional) BA. Arguing that the ratios of the energy
scales are universal (unlike the scales themselves, which do depend on the
cutoff-scheme used), they found by combining their two results (the analytical
expression is due to Hewson \cite{hew93})
\be
\xi&=&\frac{\te^{\text{C}+3/4}}{4\pi^{3/2}}=0.102676\ldots\label{anwil}
\ee
We generalize Wilson's definition \refeq{wildef} to
the general spin-$S$ case, in the presence of $m$ channels. A scheme of
numerically solving the integral equations which allows for the
calculation of the corresponding ratios is given in the appendix. For general $l$, $m$,
equation \refeq{wildef} reads in the notation of section \ref{sectgtk}
\be
2\pi \xi&=& \exp\l[-m^2\l(4\phi(m)+\frac{1}{4}\r)\r]\nn\; ,
\ee
which only depends on $m$, analogously to the ratio $\widetilde T_h/T_K$ for
$T=0$, equations \refeq{defat0}, \refeq{deftildeht0}.  This is in contradiction with
\cite{fur82}. There, the Wilson numbers for $S$ arbitrary, $m=1$ are
calculated. The ratio $T_K/T_h$ for $T=0$ is found by BA techniques for $\g=0$ and agrees
with ours, equation \refeq{thdef} for $\g=0$. $T_h/\widetilde T_K$ is found by
conventional perturbation theory. The resulting $T_K/\widetilde T_K$ depends exponentially on
$S(S+1)$. We leave this question to be clarified.  

By inserting equation \refeq{numphi}, one gets for $m=1$:
\be
\xi&=&0.102678\pm2\cdot10^{-6}\nn\; .
\ee
This result agrees with equations~\refeq{wrwil}, \refeq{anwil}. In
table \ref{tabbmfin}, Wilson numbers for the general $l\geq m$ case are given. Note
that for $2S>m$, the susceptibility at low temperatures can be obtained from
that at high temperatures by replacing $l\to(l-m)$. This does not change the
value of $\xi$, which means that in the under-screened cases, only one scale
(namely $\widetilde T_K$) governs the low- and high-temperature behavior. 
\begin{table}[h]
\begin{ruledtabular}
\begin{tabular}{cll}
\rule[-3mm]{0mm}{8mm}$m$ & $\phi(m)$   & $\xi$ \\
\hline
1 & $0.04707\pm2\cdot 10^{-5}$&$0.102678\pm2\cdot10^{-6}$\\
2 & $0.0502\pm1\cdot 10^{-4} $&$6.4604\cdot 10^{-2} \pm6\cdot10^{-6}$ \\
3 &$0.044\pm1.4\cdot 10^{-3}$& $4.455\cdot 10^{-2} \pm8\cdot10^{-5}$\\
4 &$0.038\pm1.4\cdot 10^{-3}$& $3.178\cdot 10^{-2} \pm7\cdot10^{-5}$\\
5 &$0.034\pm1.4\cdot 10^{-3}$& $2.124\cdot 10^{-2} \pm6\cdot10^{-5}$ \\
\end{tabular}
\end{ruledtabular}
\caption{\label{tabbmfin}Wilson numbers for $l\geq m$. }
\end{table}

The over-screened case $m>l$ is obtained from equations \refeq{ffasht}, \refeq{casht},
\refeq{chiasres} by replacing $m\to l$ and inserting $\ln Y_{j=l}$ and its
derivatives, equations \refeq{yj}, \refeq{yjchi}, into the definitions of the free energy, specific heat and
susceptibility. The Wilson numbers are again given in table \ref{tabbmfin}, where now $m\to l$, $\phi(m)\to\phi(l)$. In analogy to the exactly and under-screened cases, $T_K$ has been identified as a low-temperature scale also for the over-screened case in equations \refeq{ctsv}, \refeq{chitsv}. From the universality of the low-temperature Wilson ratio, we expect the Wilson number in the over-screened case also to be universal, in analogy to the exactly screened case.  

\subsection{Low temperature evaluation in the under-screened case}
\label{tlltkh0}
In this section, we make use of the results of appendix \ref{xtoinf}, from which one deduces the low-temperature behavior in the under-screened case of thefollowing quantities $(T\ll T_K)$:
\be
f_i(T)&=&\!\!
-T\ln(l-m+1)\nn\\
& &-\frac{T}{\ln^3\frac{T_K}{T}}\,\frac{m\,(l-m)(l-m+2)\pi}{12}\hspace{1cm}\nn\\
 C_i(T)&=&\!\!
\frac{T}{\ln^4\frac{T}{T_K}}\,\frac{m(l-m)(l-m+2)\pi^2}{4}\hspace{1cm}\nn\\
\chi_i(T)&=&\!\!\frac{(l-m)(l-m+2)}{12 T}\l(1+\frac{m}{\ln
  \frac{T_K}{T}}\r.\qquad\qquad\qquad\nn\\
\hspace{-1cm}& &-\l.\frac{m^2 \ln \ln T_K/T}{2\ln^2 \frac{T_K}{T}}+\frac{4m^3\phi(m)
    +m^2/4}{\ln^2\frac{T_K}{T}}\r).\label{susttun}
\ee
The low-temperature behavior for exact screening is given in
section \ref{loword}, where the expected Fermi liquid behavior shows up. Similarly
to  $T=0$, a change of sign in the leading corrections to the
asymptotic values of the susceptibility is observed,
cf. equations \refeq{chiasres}, \refeq{susttun}. Its physical interpretation has
been given in the sequel of equation \refeq{mlgeqmunil}. It applies analogously in
this case.

\subsection{Anisotropic case, $\g\neq 0$}
According to equation \refeq{restan}, the anisotropy parameter is restricted
to $0\leq \g\leq \frac{\pi}{2 n}$, where $n=\max(l,m)$. Since the kernel
$k(x)$ decays exponentially in direct space, corrections to $\ln
\B_l^{(\infty)}$, $\ln Y_j^{(\infty)}$ are expected to be exponentially
small. Thus it is no longer permitted to replace convolutions with $s(x)$ by
algebraic multiplications. Instead, let us write equation \refeq{b1} in direct
space, with the same notations as equation \refeq{not}, however including a finite
magnetic field from the beginning. First subtract the asymptotes ($\zeta:=\beta
h/2$), 
\be
\lefteqn{\lim_{x\ll 0} 2\te^{-\zeta}\frac{\sinh\zeta(l+1)}{\sinh\zeta l}\delta_l(x)= \lim_{x\ll 0}\l\{-\te^{x}\r.}\nn\\
& &+\l.\l[s*\delta_{l-1}+\kappa*\delta_{l}-\kappa_-*\ov\delta_{l}\r](x)\r\}\label{expdel}\;
.
\ee
Contrary to the isotropic case, the driving term $-\te^x$ cannot be neglected since
all quantities on the rhs of equation \refeq{expdel} are exponentially small. We
did not find a closed solution to equation \refeq{expdel}, but determine the exponent of the leading exponential decay. Therefore first note that
the equation
\be
\delta_-^{(0)}(x)&=&-\te^x+\int_{-\infty}^0 \k(x-y)\delta_-^{(0)}(y)\d y\nn
\ee 
is directly solvable by Wiener-Hopf techniques. This solution relies on the
fact that 
\be
1-\F[\k]&=& \frac{\sinh\frac{\pi k}{2}\frac\pi\g}{2\cosh\frac{\pi
    k}{2}\sinh\frac{\pi k}{2}\l(\frac\pi\g-l\r)}\nn
\ee
is factorizable in functions analytical in the upper and lower half planes. The leading
decay is $\delta_-^{(0)}(x\ll 0)\sim \te^{\frac{2\g}{\pi}x}$. We take the solution $\delta_-^{(0)}$ as an ansatz for $\delta_-$, where $\zeta$ is assumed to be small. It is seen that there is no further restriction on the leading decay of
$\delta_-$, such that
\be
\delta_-(x\ll 0)\sim \te^{\frac{2\g}{\pi}x}\nn\; .
\ee
However, we did not succeed in determining the coefficient. It
depends on $\zeta$, especially, 
\be
\!\!\!\lim_{x\ll 0}[\delta_-+\ov\delta_-](x)&\sim&(\widetilde a_1+(\beta h)^2
\widetilde a_2)\te^{\frac{2\g}{\pi}x}\nn\\
\!\!\!f_i(T\gg T_K)&=&-T\l[\ln(l+1)\r.\qquad\nn\\
\!\!\!& &\l.+(\widetilde a_1+(\beta h)^2
\widetilde a_2)\l(\frac{T_K}{T}\r)^{\frac{2\g}{\pi}}\r]\,\label{fanht}\\
\!\!\!C_i(T\gg T_K)&\sim &\l(\frac{T_K}{T}\r)^{\frac{2\g}{\pi}}\label{canht}\\
\!\!\!\chi_i(T\gg T_K)&=&\frac{l(l+2)}{12 T}\l(1- a_2 \l(\frac{T_K}{T}\r)^{\frac{2\g}{\pi}}\r).\label{chianht}\qquad 
\ee
Note that both $C_i$ and $\chi_i$ show similar decays, contrary to the isotropic
case. 

If $T\ll T_K$, one argues in close analogy to the isotropic case: The
decoupling into two independent sets still holds. However, the asymptotic values of
the auxiliary functions for
$x\to\infty$ are related to their $x\to-\infty$ counterparts by
substituting $l\to(l-m)$ {\em and} scaling $\beta h\to \alpha \beta h$,
$\alpha=(1-m\g/\pi)^{-1}$, equation \refeq{asympval}. This scaling affects the susceptibility:
\be
f_i(T\ll T_K)&=&
-T\l[\ln(l-m+1)+\r.\nn\\
& &\l.(\widetilde a_1+(\beta h\al)^2
\widetilde a_2)\l(\frac{T}{T_K}\r)^{\frac{2\g}{\pi}}\r]\,\label{fantt}\\
 C_i(T\ll T_K)&\sim &\l(\frac{T}{T_K}\r)^{\frac{2\g}{\pi}}\label{cantt}\\
 \chi_i(T\ll T_K)&=&\frac{(l-m)(l-m+2)\al^2}{12 T}\nn\\
& &\times\l(1- a_2 \l(\frac{T_K}{T}\r)^{\frac{2\g}{\pi}}\r)\label{chiantt}
\ee
The constants in equations \refeq{fanht}-\refeq{chianht} differ from those in
equations \refeq{fantt}-\refeq{chiantt}. For ease of notation, the same symbols
have been used. However, if we send $T\to 0$ first with $h\neq 0$ and $l>m$, the rigorous linearization, section \ref{lin},
is done. Then $\chi_i\sim h^{-1}\ln^{-2}h/T_K$, equation \refeq{mlgeqmunil}: The $T$-dependent power-like divergence is
replaced by a logarithmic, $h$-dependent divergence.

\section{Conclusion}
We presented a novel exact solution to the anisotropic multichannel spin-$S$
Kondo model. The free energy contribution of the impurity is given by a finite set of (max$[m,2S]$+1)-many NLIE.
By analytical and numerical studies, we confirm and extend known properties of this model.

The low temperature case is characterized by Fermi liquid behavior in the
sense that Wilson ratios are defined. However, only in the
exactly screened case $l=m$, the limits $T,h\to 0$ commute and $C_i(T)/T$,
$\chi_i$ approach finite values. These values were calculated by the dilogarithm technique and
for $l\geq m=1$ by the dressed charge formalism, in the framework of a rigorous
linearization. 

We analysed the ground state for arbitrary anisotropy, spin and channel
number and observed non-commutativity of the limits $h,T\to 0$ for models with
$l\neq m$. In the underscreened case, if $h=0$, an asymptotic
approach to free $(l-m)/2$ spin asymptotes was recovered for $T\ll T_K$, formally analogous to the $T\gg
T_K$ case, paragraph \ref{tlltkh0}. On the other hand, performing first the limit $T\to
0$ while letting $h\neq 0$, a non-integer rest spin occurs for $\g\neq
0$ in the underscreened case, equation \refeq{man}, connected to a quantum critical point \cite{sch00,sch01}. 

We analyzed the low-temperature behaviour of the over-screened models and
found non-integer exponents of $h$ if $T=0$, equation \refeq{chiovt0} and of $T$ if $h=0$, equations \refeq{chitsv}, \refeq{ctsv}. Especially,
we determined numerically low-temperature Wilson ratios for arbitrary
anisotropy, confirming and extending results by Affleck \cite{aff91}. 

At high temperatures, $T\gg T_K$, the impurity spin approaches asymptotically
the behavior of a free spin of magnitude $S=l/2$. Corrections to the
asymptotic values depend in their amplitude on the channel number
$m$ and have been calculated analytically. Especially, Wilson numbers relating low- to high-temperature scales are
determined with the help of a numerical solution. 

We expect that new insight into the multichannel case is
obtained by generalizing the lattice path integral approach proposed in
\cite{bo04} for the isotropic $S=1/2$, $m=1$ model. To this aim, $R$-matrices
must be constructed which are invariant under the action of
gl(2$|$1), containing higher dimensional irreps of su(2), \cite{sche77}.

Finally, it should be possible to derive the quantities determined numerically
in this work also by analytical methods. This question is left open for future investigations.

\appendix
\section{Asymptotic linearization}
\subsection{The region $x\to-\infty$}
\label{appa}
We begin with the asymptotic expansion of $\ln \B_l$, $\ln\BB_l$,
$l\geq m$ in the region $x\to-\infty$. The case $m>l$ is obtained therefrom
afterwards. Consider the equation for $\B_l$ in the $h=0$-case:
\be
\ln \b_l&=& s*\ln Y_{l-1}+\k*\ln \B_l-\k_-*\ln \BB_l\label{b1}
\ee  
We shall show that $\ln \B_l(x),\,\ln\BB_l(x)$ approach their asymptotes $x\to -\infty$ as $x^{-3}$ and calculate the corresponding coefficient $b_3^{(l)}$. 

The
first $l-1$ integral equations determine the $j$-dependence of $\ln Y_j(x)$,
$j=1,\ldots,l-1$. To see this, define
\be
\ln Y_j (x)&=& 2\ln (j+1) +2 \delta_j(x)\nn\\
\ln \B_l(x)&=& \ln (l+1)
+\delta_l(x)\nn \\
\lim_{x\to-\infty}\delta_j(x)&=&0\nn\\
\lim_{x\ll0}\ln y_j(x)&=& \ln j(j+2) + 2\,
\frac{(j+1)^2}{j(j+2)}\,\delta_j(x)+\O\l(\delta^2_j\r)\nn\\
\lim_{x\ll 0}\ln
\b_l(x)&=& \ln l + \frac{l+1}{l} \delta_l(x)+\O\l(\delta^2_l\r)\label{not}\; .
\ee
$\ln\BB_l$ ($\ln \bb_l$) is related to $\ln\B_l$ ($\ln\b_l$) by complex conjugation.
The crucial approximation is 
\be
[s*\delta_j](x)\approx \delta_j(x)/2 +\O(\delta_j''(x))\; .\label{crucapp}
\ee
Since $s$ is an exponentially decaying kernel, this
approximation is justified for algebraically decaying $\delta_j$. Such an
algebraic behavior is indeed expected from the integration kernel $k(x)$,
which itself decays algebraically for $\g=0$. In the anisotropic case $\g\neq
0$, this is not true, since $\delta_j(x)$ also decays exponentially. The
$\delta_j$ satisfy the recurrence relations
\begin{subequations}
\be
2\,\frac{(j+1)^2}{j(j+2)}\,\delta_j(x)&=& \delta_{j+1}(x)+\delta_{j-1}(x)\label{recac}\\
\delta_0(x)&\equiv&0\label{recb}\\
\frac{l+1}{l} \delta_l(x)&=& \delta_{l-1}(x)+3(l+1)d(x)\label{recc}\; ,
\ee
\end{subequations}
where in the last line the function $d(x)$ is to be determined and its
prefactor has been chosen by convenience. These equations determine $\delta_j$ up to a constant factor. Note that from equations \refeq{recac}, \refeq{recb}
\be
\delta_j(x) &=&  j(j+2)d(x)\nn \; .
\ee
Summarizing, 
\be
\ln Y_j(x)&=&2\ln (j+1)+2j(j+2)d(x)\nn\\
\ln \B_l(x)&=&\ln
(l+1)+l(l+1)d(x)\; .\label{yexp}
\ee 

The functions $\B_l$, $\BB_l$ are related by complex
conjugation, the $Y_i$ are real-valued. 
\be
\ln \B_l=: B_{1,l}+\ti B_{2,l} \; , \qquad \ln \BB_l=:B_{1,l}-\ti B_{2,l}\label{ccsep}\; .
\ee
Define the sum and the difference of the integration kernels,
\be
\k^{(s)}\l(x\r)&=&
\k\l(x+\ti\frac{\pi}{2}\r)+\k\l(x-\ti\frac{\pi}{2}\r)=\F^{-1}\l[\te^{-\frac{\pi}{2}|k|}\r]\nn\\
\k^{(d)}\l(x\r)&=&
\k\l(x+\ti\frac{\pi}{2}\r)-\k\l(x-\ti\frac{\pi}{2}\r)\nn\\
&=&\F^{-1}\l[\te^{-\frac{\pi}{2}|k|}\,\frac{\sinh\frac{\pi}{2}\,k}{\cosh\frac{\pi}{2}\,k}\r]\nn\; .
\ee
Asymptotically,
\be
\k^{(s)}\l(x+\ti\frac{\pi}{2}\r)&\sim&\frac{1}{2x^2}-\frac{\ti\pi}{2x^3}\;, \; |x|\to \infty\;,\nn\\
\k^{(d)}(x)&\sim&-\ti\,\frac{\pi}{2x^3}\;,\; |x|\to \infty\label{kdas}\; .
\ee
The convolutions with the $\k$-kernels are written in the following way:
\be
\k*\ln \B_l-\k_-*\ln \BB_l&=&
\k^{(d)}_-*B_{l,1}+\ti\,\k^{(s)}_-*B_{l,2}\nn\\
\k_\pm^{(\nu)}(x)&=&\k^{(\nu)}\l(x\pm\ti\frac\pi2\r)\; , \;\;\nu=s,d\nn\; .
\ee
The first non-vanishing term in an asymptotic expansion of $B_{l,1}(x)$
around $|x|\to\infty$ is 
\be
B_{l,1}(x)\sim \ln (l-m+1) + \ln \frac{l+1}{l-m+1}\theta(-x)\nn\; .
\ee
The $\theta$-function has to be understood asymptotically
for large $x$. This regime is equivalent to small $k$-values in Fourier-space,
$\F[\theta(-x)]=-\ti/k+\pi \delta(k)$. In this region around the origin in
Fourier space, $\pi k\F[\k^{(s)}]/2\sim \F[\k^{(d)}]$. Thus it follows that
\be
 \l[\k^{(d)}_-*B_{l,1}\r](x)\sim \ln
 \frac{l+1}{l-m+1}\l(\frac{\pi^2}{4x^3}-\ti\,\frac{\pi}{4x^2}\r)\label{kdbexp}\; .
\ee
It is useful to define correction terms to the asymptotic behavior for $x\ll 0$:
\be
\lim_{x\to-\infty} \ln\B_l(x)&=& \ln
(l+1)+\frac{b^{(l)}_{1,3}}{x^3}+\ti\frac{b^{(l)}_{2,2}}{x^2}\label{bcoeff}\; .
\ee
One then performs the asymptotic expansion
\be
\lefteqn{\ti \,\l[\k^{(s)}_-*B_{l,2}\r](x)}\nn\\
&\sim&\frac{\ti}{x^2}\l(\frac{1}{2}\i B_{l,2}(x)\,\d x 
+b_{2,2}^{(l)}\i \k_s(x)\,\d x\r)\label{ksbexp}\\
& &-\frac{\pi}{x^3}\l(\frac{1}{2}\i B_{l,2}(x)\,\d x
+b_{2,2}^{(l)}\,\i \k_s(x)\,\d x\r), \nn
\ee
where
\be
\i \k^{(s)}(x)\,\d x&=& 1\nn\\
\i B_{l,2}(x)\,\d x&=&\frac{\pi}{2}\l(-(m\pm 10^{-5})+\ln\frac{l+1}{l-m+1}\r)\nn\; .
\ee
The last integral is done numerically with the indicated precision. In the following, we set
\be
\i B_{l,2}(x)\,\d x&=&\frac{\pi}{2}\l(-m+\ln\frac{l+1}{l-m+1}\r)\label{intb2}\;.
\ee
Insert equation \refeq{bcoeff} in equation \refeq{not} and keep only the linear order in $\delta_l$, 
\be
\ln \b_l(x)\sim \ln l +
\frac{l+1}{l}\l(\frac{b_{1,3}^{(l)}}{x^3}+\ti\,\frac{b_{2,2}^{(l)}}{x^2}\r)\; .\label{lnbexp}
\ee
Combining equations \refeq{kdbexp}, \refeq{ksbexp}, \refeq{lnbexp}
one expands equation \refeq{b1} around $x\to -\infty$:
\be
\lefteqn{\ln l + \frac{l+1}{l} \,\frac{b_{1,3}^{(l)}}{x^3}= \ln l + (l-1)(l+1)d(x)}\nn\\
& &+\frac{\pi^2}{4x^3}\,\ln
    \frac{l+1}{l-m+1} -\frac{\pi}{x^3}\l(\frac12\i B_{2,l}(x)\d
    x+b_{2,2}^{(l)}\r)\nn\\
\lefteqn{\frac{l+1}{l}
\,\frac{b_{2,2}^{(l)}}{x^2}=\frac{1}{x^2}\l(-\frac\pi4\ln\frac{l+1}{l-m+1}\r.}
\nn\\
& &\l.+\frac12
\i B_{2,l}(x)\d x+b_{2,2}^{(l)}\r)\nn\; .
\ee
Using equation \refeq{intb2}, we find:
\be
d(x)&=& \frac{\pi^2}{12}\, \frac{m}{x^3}\;,\;
b_{2,2}^{(l)}= -\frac{l\,m\,\pi}{4}\nn\\
b_{1,3}^{(l)}&=& \frac{l(l+2)\,m}{3}\,\frac{\pi^2}{4}\nn\;,\\
\lim_{x\to-\infty} \ln Y_j(x)&=&2\ln (j+1)+j(j+2)\,\frac{m\pi^2}{6x^3}\label{yj},\\
\lim_{x\to-\infty}\ln \B_l(x)&=& \ln (l+1) -\ti
\frac{m\,l\,\pi}{4\,x^2}+\frac{m\,l(l+2)\pi^2}{12\,x^3}\nn
\; .
\ee
Note that the $x$-dependence of the corrections is determined through
the asymptotic behavior of the kernel in the convolutions $\k*\ln \B_l$,
$\k*\ln\BB_l$. The amplitudes follow from the $Y$-hierarchy. 

We proceed with the asymptotic evaluation of $\partial_{\beta h} \ln
Y_j(x)$, $\partial_{\beta h} \ln \B_l(x)$ in the regime $x\ll 0$. The
$\ln Y_j$ are symmetric with respect to $\beta h$: The system of NLIE remains
the same upon replacing $\beta h\to-\beta h$ and substituting $\B_l$ by $\BB_l$. So
$\partial_{\beta h}\ln Y_j$ vanishes identically for $h=0$. Define
$B_l^{(m)}:=\partial_{\beta h} \ln \B_l$. Then the only equation which remains
is
\be
B_l^{(m)}(x)&=& \l(1-\te^{-\ln \B_l(x)}\r)\l[\frac{1}{2}+[\k*B_l^{(m)}\r.\nn\\
& &\l. - \k_-*\ov B_l^{(m)}](x)\r]\label{bma}\; .
\ee
$B_l^{(m)}$ and $\ov B_l^{(m)}$ are related by change of sign and complex
conjugation,
\be
B_l^{(m)}&=:& B^{(m)}_{l,1}+\ti B^{(m)}_{l,2} \nn\\
 \ov B_l^{(m)}&=&-\l[B_l^{(m)}\r]^*=
-B^{(m)}_{l,1}+\ti B^{(m)}_{l,2}\nn\; .
\ee 
Below it is shown that the imaginary part of $B_l^{(m)}$
vanishes for $h=0$, so $B_l^{(m)}\equiv
B^{(m)}_{l,1}$ is real valued. We shall determine asymptotic
corrections to $B_l^{(m)}$ up to the order $\O\l(x^{-2}\r)$, so that corrections to
$\ln \B_l(\pm \infty)$ (of order $\O\l(x^{-3}\r)$), can safely be neglected. In
this approximation, 
\be
\lim_{x\to -\infty} \l(1-\te^{-\ln \B_l(x)}\r)\!\!&=&\!\!\frac{l}{l+1}\l(1+\ti\frac{\pi}{4x^2}\r)\nn
\ee
\be
 \lim_{x\to \infty}
\l(1-\te^{-\ln \B_l(x)}\r)\!\!&=&\!\!\frac{l-m}{l-m+1}\l(1+\ti\frac{\pi}{4x^2}\r) ,
\ee
and one finds the asymptotic equation for $B^{(m)}_l$,
\be
\lim_{x\ll 0} B^{(m)}_l(x)&=&\frac{l}{l+1} \l(\frac12+\k^{(s)}*B_l^{(m)}\r)\label{comlin}\;.
\ee
The imaginary contributions vanish as expected. An equation analogous to \refeq{comlin} does not exist in the TBA-approach, where one deals with
the infinitely many $Y_j$ and their derivatives. But $Y_j^{(m)}\equiv 0 \, \forall j$, as mentioned above.  

The constant asymptotic behavior of $B^{(m)}_l$ is written in the compact form
\be
B_l^{(m)}(x)&=& \Delta_+-\Delta_-\sgn x\label{comp}\\
\Delta_++\Delta_-&=&\frac{l}{2} \qquad \Delta_+-\Delta_-=\frac{l-m}{2}\nn\; .
\ee
Equation \refeq{comp} and similar equations in the following have to be understood
asymptotically, for $|x|\to\infty$. 
In a similar manner, 
\begin{subequations}
\be
\k^{(s)}*\sgn x&=&\sgn x -\frac{1}{ x}\nn\\
  \k^{(s)}*\frac{\sgn x}{x}
&=&\frac{\sgn x}{ x} + \frac{\ln |x|}{ x^2} - \frac{\Psi(2)}{
  x^2}\label{kconv}\\
\k^{(s)}*\frac{1}{x}&=& \frac{1}{ x}\,,\qquad
\k^{(s)}*\frac{\ln |x|}{x^2}=\frac{\ln |x|}{x^2}\label{kconvb}
\; .
\ee
\end{subequations}
$\Psi(x)$ is the digamma function; $\Psi(2)=-\text C +1$, $\text
C=0.577\ldots$ is Euler's constant. The asymptotic evaluation of the
convolutions is done by using
distributions. To determine accurately
the $x^{-2}$-coefficient, the precise behavior of the auxiliary functions
around the origin must be known. This is done numerically as shown in the
appendix. 

We make the following ansatz for $B^{(m)}_l$, which consists
in extrapolating the asymptotic behavior over the whole axis with the aid of distributions:
\begin{widetext}
\be
B^{(m)}_l(x)&=& \Delta_+-\Delta_-\sgn
x+\frac{1}{x}(\Delta_+^{(1)}-\Delta_-^{(1)}\sgn x) + \frac{\ln |x|}{x^2}
(\Delta_+^{(1)} - \Delta_-^{(1)}\sgn x)+\frac{1}{x^2}(\Delta_+^{(2)} - \Delta_-^{(2)}\sgn x)\hspace{1cm}\label{ansbm}\\
\l[\k*B^{(m)}_l\r](x)&=&\frac{\Delta_+-\Delta_-\sgn x}{2} + \frac{1}{2 x}
\l(\Delta_-+ \Delta_+^{(1)}-\Delta_-^{(1)}\, \sgn x\r)-\l(\Delta_-^{(1)} -\Delta_+^{(l)}+\Delta_-^{(l)}\sgn x\r)\, \frac{\ln
  |x|}{2x^2}\nn\\
& &+ \l(\Delta_-^{(1)}\Psi(2)+\Delta_+^{(2)}-\Delta_-^{(2)}\sgn x+\frac{\varepsilon}{2}\r)\,\frac{1}{2x^2}\hspace{1cm}\label{convas}\; ,
\ee
\end{widetext}
where $\varepsilon$ is determined numerically (cf. appendix, equation \refeq{numeps}). The convolutions in the second equation are done with the help of equations \refeq{kconv},
\refeq{kconvb}. The terms $\sgn x$ and $\sgn x/x$ give the next higher
order contribution when convoluted with the kernel. The
asymptotic form of equation \refeq{comlin} is found by inserting equations \refeq{ansbm}, \refeq{convas}.
By comparing coefficients, one finds
\be
\Delta_+^{(1)}+\Delta_-^{(1)}&=&l\,\Delta_-=\frac{m\cdot l}{4}\qquad
\Delta_+^{(1)}=\frac{m(2l-m)}{8} \nn\\
\Delta_+^{(1)}-\Delta_-^{(1)}&=&(l-m)\Delta_-=\frac{m(l-m)}{4}\qquad \Delta_-^{(1)}=\frac{m^2}{8}\nn\\
\Delta_+^{(l)}+\Delta_-^{(l)}&=&-l\Delta^{(1)}_-=-\frac{m^2\cdot l}{8}\nn\\
\Delta_+^{(l)}-\Delta_-^{(l)}&=&-(l-m)\Delta^{(1)}_-=-\frac{m^2(l-m)}{8}\nn\\
\Delta_+^{(2)}+\Delta_-^{(2)}&=&l(\Delta^{(1)}_-\Psi(2)+\varepsilon/2)\label{del22p}\\
\Delta_+^{(2)}-\Delta_-^{(2)}&=&(l-m)(\Delta^{(1)}_-\Psi(2)+\varepsilon/2)\label{del22m}
\ee
In the high-temperature regime, the $x\to-\ln T/T_K\ll 0$ behavior is of importance,
\be
\lim_{x\ll 0}B_l^{(m)}(x)&=& \frac{l}{2} + \frac{m \,l}{4\,x}-\frac{m^2\,
  l}{8}\, \frac{\ln |x|}{x^2}\nn\\
& &+\frac{l}{x^2}\l(\frac{m^2\,\,\Psi(2)}{8}+\frac{\varepsilon}{2}\r)\label{maght}\; .
\ee
Compare equation \refeq{maght} with the low temperature results
equations \refeq{mlgeqmunih}, \refeq{mlgeqmunil}.  Both are formally identical up to
the order $\ln|x|/x^2$, $x=-\ln T/T_K$ in equation \refeq{maght}, $x=-\ln h/\widetilde T_h$ in
equations \refeq{mlgeqmunih}, \refeq{mlgeqmunil}. 

We shall focus on the $x\to+\infty$ asymptotes in the next paragraph. 

Our analysis is continued by expanding $\partial^2_{\beta h}\ln Y_j=:Y_j^{(\chi)}$,
$\partial^2_{\beta h} \ln \B_l=:B_l^{(\chi)}$, $\partial^2_{\beta h} \ln
\BB_l=:\ov B_l^{(\chi)}$. These functions are given by the system
\begin{subequations}
\be
Y_j^{(\chi)}&=&\!\!\! \l(1-\te^{-\ln Y_j}\r)s*\l(Y_{j-1}^{(\chi)}+Y_{j+1}^{(\chi)}\r)\hspace{0.8cm}\label{chia} \\
Y_0^{(\chi)}&\equiv&\!\!\!0\hspace{0.8cm}\label{chib}\\
B_l^{(\chi)}&=&\!\!\! \frac{\te^{-\ln \B_l}}{1-\te^{-\ln \B_l}}
\l[B_l^{(m)}\r]^2 + \l(1-\te^{-\ln
  \B_l}\r) s*Y_{l-1}^{(\chi)} \nn\\
& &+ \l(1-\te^{-\ln \B_l}\r) \l(\k*B_l^{(\chi)}-\k_-*\ov
B_l^{(\chi)}\r)\hspace{0.8cm}\label{chic}
\ee
\be
B_l^{(\chi)}&=:&\!\!\! B_{l,1}^{(\chi)}+\ti B_{l,2}^{(\chi)}\hspace{0.8cm}\nn\; .
\ee
\end{subequations}
The vanishing of $B_2^{(m)}$ helps to
expand $B_1^{(\chi)}$:
\begin{subequations}
\be
\lefteqn{B^{(\chi)}_{l,1}=\frac{1}{l}\l[B_{l,1}^{(m)}\r]^2+\frac{l}{2(l+1)} \, Y_{l-1}^{(\chi)}}\nn\\
& &+
\frac{l}{l+1}\, \re\l(\k_-^{(d)}*B_{l,1}^{(\chi)}+\ti \k_-^{(s)}*B_{l,2}^{(\chi)}\r)\label{e1}\\ 
\lefteqn{B^{(\chi)}_{l,2}=-\frac{\pi m}{4x^2}\l[B_{l,1}^{(m)}\r]^2+\frac{lm}{l+1}\frac{\pi}{8x^2}\,Y_{l-1}^{(\chi)}}\nn\\
& &+
\frac{l}{l+1}\im\l( \k_-^{(d)}*B_{l,1}^{(\chi)}+\ti \k_-^{(s)}*B_{l,2}^{(\chi)}\r)  \label{e2} .
\ee
\end{subequations}
From equation \refeq{e2}, $B^{(\chi)}_{l,2} = \O\l(x^{-2}\r)$. Together with
equation \refeq{kdas}, one concludes that the last term in brackets in equation \refeq{e1} is $\O\l(x^{-3}\r)$. This means that for our purposes, the
convolutions in equation \refeq{e1} can be entirely neglected. Thus in the
asymptotic limit,
equations \refeq{chia}-\refeq{chic} are simplified considerably:
\be
Y_j^{(\chi)}&=&
\frac{j(j+2)}{2(j+1)^2}\l(Y_{j-1}^{(\chi)}+Y_{j+1}^{(\chi)}\r)\nn\\
Y_0^{(\chi)}&\equiv&0\nn\\
B_l^{(\chi)}&=&\frac{1}{l} \,\l[B_l^{(m)}\r]^2+\frac{l}{2(l+1)}Y_{l-1}^{(\chi)}\nn
\ee
This system bears similarity with equations \refeq{recac}-\refeq{recc}. From the
solution of those equations, we conclude
\be
\lefteqn{\lim_{x\to -\infty}Y_j^{(\chi)}(x)= \frac{j(j+2)}{6}\,\l(1+\frac{m}{x} -\frac{m^2\,\ln |x|}{2
x^2}\r.}\nn\\
& & \l.+\l(\frac{m^2}{2}\Psi(2)+\frac{m^2}{4}+2\varepsilon\r)\frac{1}{x^2}\r)\label{yjchi}\\
\lefteqn{\lim_{x\to -\infty}B_l^{(\chi)}(x)=
b_{l,0}^{(\chi)}+\frac{b_{l,1}^{(\chi)}}{x} + b_{l,2}^{(\chi)}\frac{\ln |x|}{x^2}+\frac{b_{l,2}^{(\chi)}}{x^2}}\nn\\
&=&\frac{l(l+2)}{12}\, \l(1+\frac{m}{x}-\frac{m^2\, \ln |x|}{2
x^2}\r.\nn\\
&& \l.+\l(\frac{m^2}{2}\Psi(2)+\frac{m^2}{4}+2\varepsilon\r)\frac{1}{x^2}\r)\nn
\ee
In the appendix, a procedure for
numerically determining the coefficient $\Delta^{(2)}_+-\Delta^{(2)}_-\sgn x$
of the $x^{-2}$-decay is described. It deviates slightly from the analytical
estimate. We introduce an extra symbol $\phi$,
\be
\phi(m):=\frac{\l(\Delta^{(2)}_++\Delta^{(2)}_-\r)}{lm^3}=\frac{1}{m^3}\l(\frac{m^2}{8} \Psi(2)+\frac\varepsilon2\r)\label{phidef}
\ee
Numerically, $\phi$ is found to be independent of $l$. However, our data do
not suffice to exclude a dependence on $m$. The results are given in table \ref{tabbmfin} in the main body of this work.

If $l<m$, one replaces in the whole preceding analysis of this appendix $l\to m$ and $m\to l$. The thermodynamical quantities are given by $Y_l$ and its derivatives. 

\subsection{The region $x\to\infty$}
\label{xtoinf}
We determine the behavior of the auxiliary functions for large values
$x\to \infty$ in the under-screened case $l>m$. In this limit, the set of NLIE decouples in two
separate sets with exponential accuracy. Consider first the field-free case $h=0$. The last $l-m$ functions,
namely $Y_{m+1},\ldots,\B_{l},\BB_{l}$, satisfy a set of equations
formally identical to equations \refeq{recac}-\refeq{recc}, with $j,l$ replaced by
$j-m,l-m$. The whole analysis of the preceding paragraph \ref{appa} applies to this case;
of special interest are now the corrections to the $x\sim\ln \frac{T_K}{T}\gg
0$-behavior of the auxiliary functions. The results are:
\be
\B_l(x)&=& \ln (l-m+1) -\ti
\frac{m\,(l-m)\,\pi}{4\,x^2}\nn\\
& &+\frac{m\,(l-m)(l-m+2)\pi^2}{12\,x^3}\nn\\
\ln Y_{j>m}(x)&=& \ln
(l+1-j)+\frac{m\,(l-m)(l-m+2)\pi^2}{6\,x^3}\nn\\
 B_l^{(m)}(x)&=&\frac{l-m}{2} + \frac{m \,(l-m)}{4\,x}-\frac{m^2\,
  (l-m)}{8}\, \frac{\ln |x|}{x^2}\nn\\
& &+\frac{l-m}{x^2} \, \l(\frac{m^2\Psi(2)}{8}+\frac{\varepsilon}{2}\r)\nn\\
Y_j^{(\chi)}(x)&=& \frac{(j-m)(j-m+2)}{6}\,\l(1+\frac{m}{x} -\frac{m^2\,\ln |x|}{2
x^2}\r.\nn\\
& &+\l.\l(\frac{m^2}{2}\Psi(2)+\frac{m^2}{4}+2\varepsilon\r)\frac{1}{x^2}\r)\nn\\
B_l^{(\chi)}(x)&=& \frac{(l-m)(l-m+2)}{12}\, \l(1+\frac{m}{x}-\frac{m^2\, \ln |x|}{2
x^2}\r.\nn\\
& &+\l.\l(\frac{m^2}{2}\Psi(2)+\frac{m^2}{4}+2\varepsilon\r)\frac{1}{x^2}\r)\nn
\ee

\section{Details of the numerical treatment}
From the analytical analysis of the
NLIE, numerically ill-conditioned terms were found, namely those $\O(1)$,
$\O(x^{-1})$ in the limit $|x|\to\infty$.\footnote{Algebraic corrections to
  the asymptotes only occur in the isotropic case.} Since in Fourier space,
they would appear as simple poles and discontinuities in the origin, they are subtracted from the
auxiliary functions {\em before} performing the FFT and treated separately. After having applied
the inverse FFT, the analytically convoluted terms are re-added:
\be
\k* \ln Y&=& \underbrace{\k*(
  \ln Y-f)}_{\text{N}}+\underbrace{\k*f}_{\text{A}}\label{v2}\;,\\
\lim_{|x|\to\infty} \ln Y(x)&\sim& f(x) \label{aseq}\; ,
\ee
and $\k$ is a kernel, $\ln Y$ an auxiliary function. The function $f$ is chosen to be asymptotically equal to
$\ln Y$, equation \refeq{aseq} and contains the numerically ill-conditioned contributions. The difference $\ln Y -f$ is therefore numerically transformable. The term denoted by N is treated numerically,
that labeled by A analytically. The Fourier-transforms of $f$ and $k$ are
known. The convolution $k*f$ is solved separately, such
that no further manipulations in Fourier space are necessary. We solve numerically for $(\ln Y-f)$: By including terms $\O(1)$,
$\O(x^{-1})$ and $\O(\ln|x|/x^2)$ in $f(x)$, the function $(\ln Y-f)$ decays as
$x^{-2}$. Since the kernels also decay as $x^{-2}$, the coefficient of the
$x^{-2}$ decay of $\ln Y$ is obtained from the integral of $(\ln Y-f)$, a
quantity which is accessible numerically with high accuracy. Consider the case
$\ln Y\equiv  B_l^{(m)}$, given by \refeq{bma}, and define the regular
function $B_{reg}:=B_l^{(m)}-f$, such that $B_{reg}(|x|\to\infty)\sim
x^{-2}$. Then for $|x|\to\infty$
\be
\k*B_{reg}(x)&\sim& \k(x)\,\i B_{reg}(x)\,\d x \nn\\
& &+ B_{reg}(x)\,\i \k(x)\,\d x\label{numeps}\;,
\ee
and the quantity $\varepsilon$ in equation \refeq{convas} is identified as
$\varepsilon:= \i B_{reg}(x)\,\d x$.

\end{document}